\documentclass[pre,longbibliography]{revtex4-1}
\usepackage{amsmath,amssymb,latexsym,epsfig,graphics,epsf}
\usepackage{graphics,xcolor}
\usepackage{float}
\graphicspath{{figures/}}
\newcommand{\br}{\bold r}

\newcommand{\bR}{\bold R}
\newcommand{\tbR}{\tilde{\bold R}}
\newcommand{\be}{\begin{equation}}
\newcommand{\ee}{\end{equation}}
\newcommand{\fig}[1]{Fig.~\ref{#1}}
\newcommand{\Fig}[1]{Figure~\ref{#1}}
\newcommand{\Sec}[1]{Sec.~\ref{#1}}
\newcommand{\eq}[1]{Eq.~(\ref{#1})}
\newcommand{\Eq}[1]{Equation~(\ref{#1})}
\newcommand{\Sex}{{S}_{\rm ex}}
\newcommand{\bRa}{{\bf R}_{\rm a}}
\newcommand{\bRb}{{\bf R}_{\rm b}}

\begin{document}
\title{Effectively one-dimensional phase diagram of CuZr liquids and glasses}
\author{Laura Friedeheim}
\author{Nicholas P. Bailey}
\author{Jeppe C. Dyre}
\email{dyre@ruc.dk}
\affiliation{``Glass and Time'', IMFUFA, Dept. of Science and Environment, Roskilde University, P. O. Box 260, DK-4000 Roskilde, Denmark}
\date{\today}

\begin{abstract} 
This paper presents computer simulations of Cu$_x$Zr$_{100-x}$ $(x=36,50,64)$ in the liquid and glass phases. The simulations are based on the effective-medium theory (EMT) potentials. We find good invariance of both structure and dynamics in reduced units along the isomorphs of the systems. The state points studied involve a density variation of almost a factor of two and temperatures going from 1500 K to above 4000 K for the liquids and from 500 K to above 1500 K for the glasses. For comparison, results are presented also for similar temperature variations along isochores, showing little invariance. In general for a binary system the phase diagram has three axes: composition, temperature and pressure (or density). When isomorphs are present, there are effectively only two axes, and for a fixed composition just one. We conclude that the liquid and glass parts of the thermodynamic phase diagram of this metallic glass former at a fixed composition is effectively one-dimensional in the sense that many physical properties are invariant along the same curves, implying that in order to investigate the phase diagram, it is only necessary to go across these curves.
\end{abstract}
\maketitle

\section{Introduction}\label{Intro}

Metallic systems constitute a very important category of glass formers due to their potential applications, as well as their suitability as model systems for studies of the glass transition in computer simulations \cite{gre95,zin06,men09,wan12,wan19}. A well-studied example is the CuZr system, which is a fairly good glass former despite consisting of just two elements \cite{tan04,men09}. This paper presents numerical evidence that both above and below the glass transition, CuZr systems are simpler than has hitherto been recognized. Specifically, for three different compositions of the CuZr system we show that curves exist in the thermodynamic phase diagram along which the atomic structure and dynamics are invariant to a good approximation. The implication is that the two-dimensional thermodynamic phase diagram becomes effectively one-dimensional in regard to many material properties.

The background of the investigation is the following. In liquid-state theory, a simple liquid is traditionally defined as a single-component system of particles described by classical Newtonian mechanics and interacting by pair-potential forces \cite{ric65,tem68,sti75,barrat,ing12,han13}. It has been known for more than half a century that the hard-sphere (HS) model reproduces well the physics of many simple liquids, both in regard to the radial distribution function (RDF) and to dynamic properties such as the viscosity or the diffusion coefficient \cite{moltheory,ber64,wid67,bar76,cha83,han13,dyr16}. The traditional explanation of this is the ``van der Waals picture'' according to which the repulsive forces dominate the physics of simple liquids \cite{vdW1873,wid67,bar76,cha83,han13}. 

The HS model is a caricature simple liquid with pair forces that are zero except right at the particle collisions. In the HS model, temperature plays only the trivial role of determining particle velocities and thus the time scale; temperature is entirely unrelated to the geometry of particle positions. This implies that the thermodynamic phase diagram of the HS system is effectively one-dimensional with density as the only non-trivial variable: the dynamics of two different HS systems with the same packing fraction but different temperatures are identical, except for a trivial uniform scaling of the space and time coordinates. As a consequence, scaled RDFs are identical, scaled mean-square displacements are identical, viscosities are trivially related, etc.

The mapping of a simple liquid to a HS system presents the issue of identifying the effective HS packing fraction at a given thermodynamic state point of the liquid. Many suggestions have been made for how to calculate the relevant hard-sphere radius, yet no consensus was ever arrived at \cite{row64,hen70,and71,kan85,har92, str92,ben04,nas08,rod13}. Already in 1977, Rosenfeld suggested a powerful thermodynamics-based alternative by basically reasoning as follows \cite{ros77}: Since the HS packing fraction determines the configurational part of the entropy, this quantity provides the required mapping between a simple liquid and its corresponding HS system. Defining the excess entropy $\Sex$ as the system's entropy minus that of an ideal gas at the same density and temperature \cite{ros77,han13}, $\Sex$ quantifies the configurational entropy (note that $\Sex<0$ because any system is less disordered than an ideal gas). Rosenfeld's suggestion implies invariance of the physics along the curves of constant excess entropy in the phase diagram. He validated this from that time's fairly primitive computer simulations of the Lennard-Jones system and a few other simple liquids \cite{ros77}. Rosenfeld's insight is now referred to as ``excess-entropy scaling'', a property that has received increasing attention since the turn of the century because it has been found to apply also for many non-simple systems like liquid mixtures, molecular liquids, confined liquids, crystalline solids, etc \cite{dyr18a}. 

In the same time period of the last 20 years, glass science has progressed significantly by the introduction of density scaling, also called thermodynamic scaling. This is the discovery that in the search for a simple mathematical description, the relevant thermodynamic variables are not temperature and pressure, but temperature $T$ and the particle number density $\rho$ \cite{kiv96,alb04,rol05,gun11,lop12a,adr16}. When density scaling is applied to experimental data, if $\gamma$ is the so-called density-scaling exponent, plotting data for the dynamics as a function of $\rho^\gamma/T$ results in a collapse \cite{alb04,rol05,lop12a,adr16}. This means that the dynamics depends on the two variables of the thermodynamic phase diagram only via the single number $\rho^\gamma/T$. It should be emphasized that density scaling is not universally applicable;  for instance, it works better for van der Waals liquids and metals than for hydrogen-bonded liquids \cite{rol05,adr16,hu16}. An important extension of density scaling was the discovery of isochronal superposition, according to which not only the average relaxation time is invariant along the curves of constant $\rho^\gamma/T$, so are the frequency-dependent response functions \cite{rol03,nga05,han18}. This indicates that the way atoms or molecules move about each other is identical at state points with same the same value of $\rho^\gamma/T$. One may think of this as a ``same-movie'' property: Filming the atoms/molecules at two such state points results in the same movie except for a uniform scaling of all particle positions and of the time.

The above-mentioned findings can all be derived from the hidden-scale-invariance property stating that the ordering of a system's configurations $\bR\equiv (\br_1,...,\br_N)$ (in which $N$ is the number of particles and $\br_i$ is the position of particle $i$) according to their potential energy, $U(\bR)$, at one density is maintained if the configurations are scaled uniformly to a different density \cite{sch14}. The formal mathematical definition of hidden scale invariance is the following logical implication

\be\label{hsi}
U(\bRa)<U(\bRb)\,\,\Rightarrow\,\, U(\lambda\bRa)<U(\lambda\bRb)\,.
\ee
\Eq{hsi} implies that structure and dynamics, when given in proper reduced units, are invariant along the curves of constant excess entropy, the system's so-called isomorphs \cite{IV,sch14,dyr18a}. This result is rigorous if \eq{hsi} applies without exception, but this is never the case for realistic models. However, isomorph invariance is still a good approximation if  \eq{hsi} applies for most of the physically important configurations and for scaling parameters $\lambda$ relatively close to unity. This is believed to be the case for many metals and van der Waals bonded systems, whereas systems with strong directional bonds like hydrogen-bonded and covalently bonded systems are not expected to obey isomorph-theory predictions \cite{dyr14}. For metals, however, the existence of isomorphs has only been validated in a few cases \cite{hum15,fri19}.

In isomorph theory the density-scaling exponent $\gamma$ is generally state-point dependent. This has recently been confirmed in high-pressure experimental data \cite{cas19,san19,cas20}. Systems with hidden scale invariance are referred to as R-simple in order to indicate the simplification of the physics that follows from this symmetry; most, though not all, pair-potential systems are R-simple and several molecular systems are also R-simple, necessitating a specific name for this class of systems.

The purpose of the present paper is to check for isomorphs in a typical metallic glass former. For this we have chosen to study three different CuZr mixtures. As a representative of the Cu-rich alloys that have been most commonly studied in experiments \cite{wan12,wan19,tan04}, we have chosen the 64:36 composition. Supplementing this, we also simulated the 50:50 and the 36:64 compositions. The findings of all three systems are similar. The systems have been computer simulated both in the liquid and glass phases, using the effective-medium theory (EMT) interaction potential \cite{pus81,nor82,jac96}. We find good isomorph invariance of structure and dynamics involving density changes up to a factor of two. This implies a significant simplification in the description of the physics of this metallic glass former since the thermodynamic phase of diagram of CuZr is effectively one-dimensional.

\section{The Effective-Medium-Theory (EMT) potential}\label{EMT_sec}

The EMT potential \cite{pus81,nor82,jac96} is one of several similar potentials aimed at describing metals with an accuracy comparable to that of a full density-functional-theory (DFT) treatment, but at a much lower computational cost. A widely used class of potentials in this group of mean-field potentials is the embedded atom method (EAM) \cite{foi86,daw93}. EMT and EAM  both write the total energy $E$ as a pair-potential part plus a function of the local electron density at each particle. The EMT realizes this in a semi-empirical way, whereas the parameters of the EAM are determined by fitting to experimental properties of the bulk solid. For more on the relation between the EAM and EMT potentials, the reader is referred to Refs. \onlinecite{pus81,nor82}, while Ref. \onlinecite{jac96} gives a detailed derivation of the EMT potential and its parameters. A great advantage of the EMT is that the mathematical expression for the energy is relatively simple. This made it straightforward to implement EMT in our GPU-code RUMD \cite{RUMD}, whereas the EAM typically involves tabulated data that are not easily implemented efficiently in GPU computing.

The core of the EMT potential is a well chosen reference system defining the \textit{effective medium}. The total energy of the system is the energy of the reference system plus the difference to the real system. Thus the EMT total energy is written 

\be
E = \sum_i E_{c,i} + \left(E-\sum_i E_{c,i} \right)
\ee
where $E_{c,i}$ is the so-called cohesive energy, which is the energy of atom $i$ in the reference system. The idea is now that the difference term should be small enough to be treated accurately by first-order perturbation theory. To obtain this the reference system must be as close as possible to the real system. 

The real and the reference systems are linked by a ``tuning parameter''. In the first version of the EMT potential, the homogeneous electron gas was used as the reference system, with the electron density as the tuning parameter \cite{jac87,nor80}. The EMT version used in this paper is that of Ref. \onlinecite{jac96} for which a perfect fcc crystal is the reference system. Here, the lattice constant serves as the tuning parameter, i.e., is used to adjust the environment of an atom such that the average electron density surrounding the atom matches that of the real system. 

The EMT potential of a pure metal involves seven parameters: the negative cohesive energy $E_0$, a charge density parameter $n_0$, the Wigner-Seitz radius $s_0$ (defined in terms of the atomic, not the electronic density), a parameter $\eta$ quantifying the influence of the density tail surrounding an neighboring atom, a parameter $\lambda$ determined from the bulk modulus, and finally the product $V_0 (\beta \eta_2- \kappa)$ determined from the shear modulus. The density-related parameters $n_0$ and $\eta$ were originally calculated self-consistently by reference to the homogeneous electron gas, while the five other parameters are determined from experimental or \textit{ab initio} data. More details on how the parameters are determined for single and compound systems can be found in in Refs. \onlinecite{jac87,pad07}. The parameter values for CuZr used in this work (Table \ref{table1}) are those of Ref. \onlinecite{pad07}, where parameters were adjusted to match DFT-determined properties—cohesive energies, lattice constants, and elastic constants of both the pure metals and a Cu$_{50}$Zr$_{50}$ alloy structure.

\begin{table}[b!]
	\centering
	\begin{tabular}{c|c c c c c c c}
		Material & $s_0$ (\AA) & $E_0$ (eV) & $\lambda$ (\AA$^{-1}$) & $\kappa$ (\AA$^{-1}$) & $V_0$ (eV) & $n_0$ (\AA$^{-3}$) & $n_2$ (\AA$^{-1}$) 
		\\ \hline
		Cu & 1.41 & -3.51 & 3.693 & 4.943 & 1.993 & 0.0637 & 3.039
		\\
		Zr & 1.78 & -6.30 & 2.247 & 3.911 & 2.32 & 0.031 & 2.282
	\end{tabular}
	\caption{EMT parameters for Cu and Zr in CuZr mixtures \cite{pad07}.}
	\label{table1}
\end{table}

\section{Isomorph theory}

\subsection{Reduced quantities}\label{reduced}

Structure and dynamics are invariant along isomorphs only when these are given in so-called reduced versions based on a ``macroscopic'' unit system that depends on the state point in question. The unit system defining reduced variables reflects the system's volume $V$ and temperature $T$ as follows. If the particle number density is $\rho\equiv N/V$, the length, energy, and time units are, respectively, \cite{IV}

\be\label{red_units}
l_0=\rho^{-1/3}\,\,,\,\,e_0=k_BT\,\,,\,\,t_0=\rho^{-1/3}\sqrt{\frac{m}{k_BT}}\,.
\ee
Here $m$ is the average particle mass. \Eq{red_units} refers to Newtonian dynamics; for Brownian dynamics one uses the same length and energy units, but a different time unit \cite{IV}. All physical quantities can be made dimensionless by reference to the above units. ``Reduced'' quantities are denoted by a tilde, for instance

\be\label{tbReq}
\tbR
\,\equiv\,\rho^{1/3}\bR\,.
\ee

\begin{figure}[t]
	\includegraphics[width=5cm]{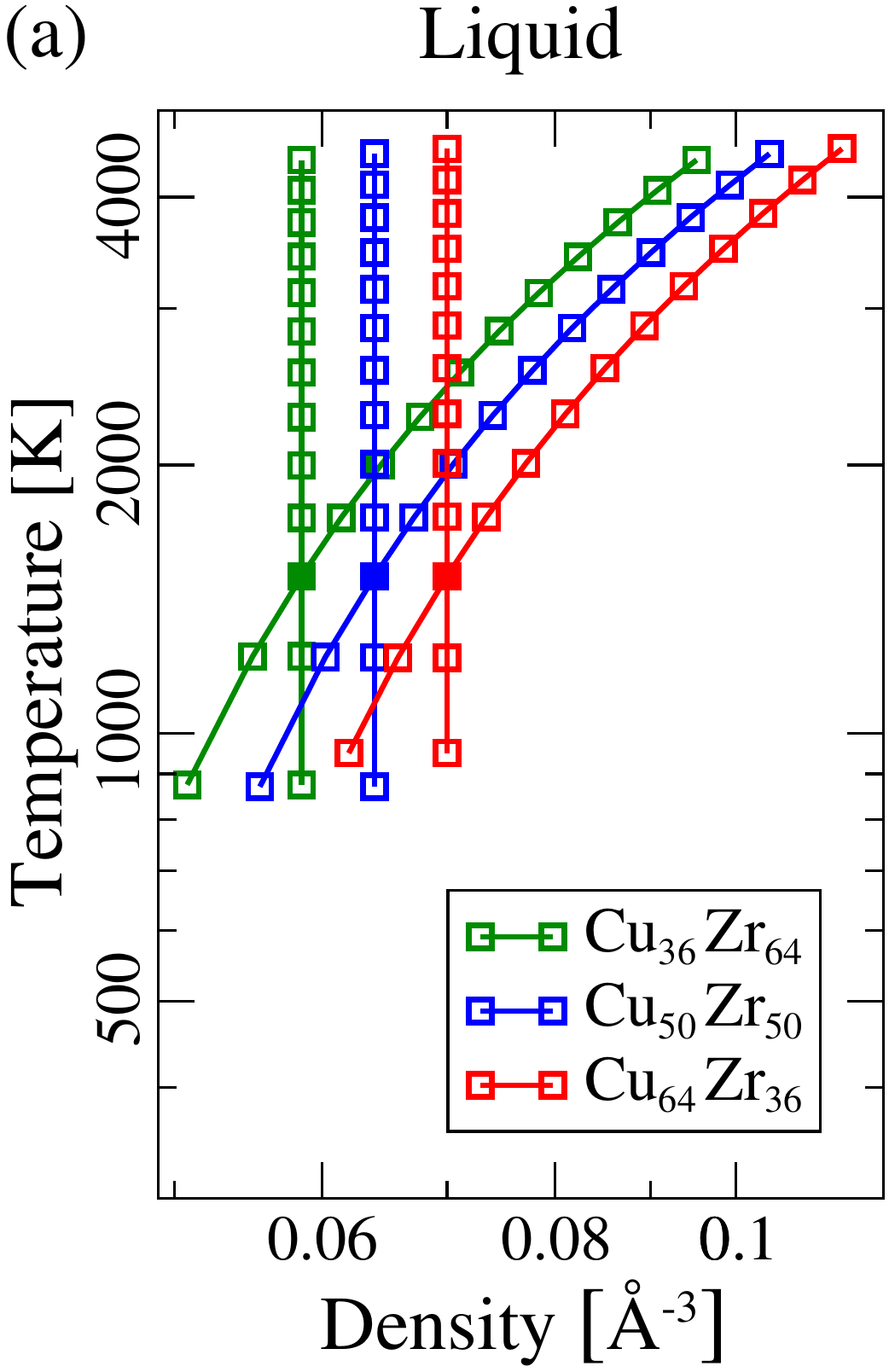}
	\includegraphics[width=5cm]{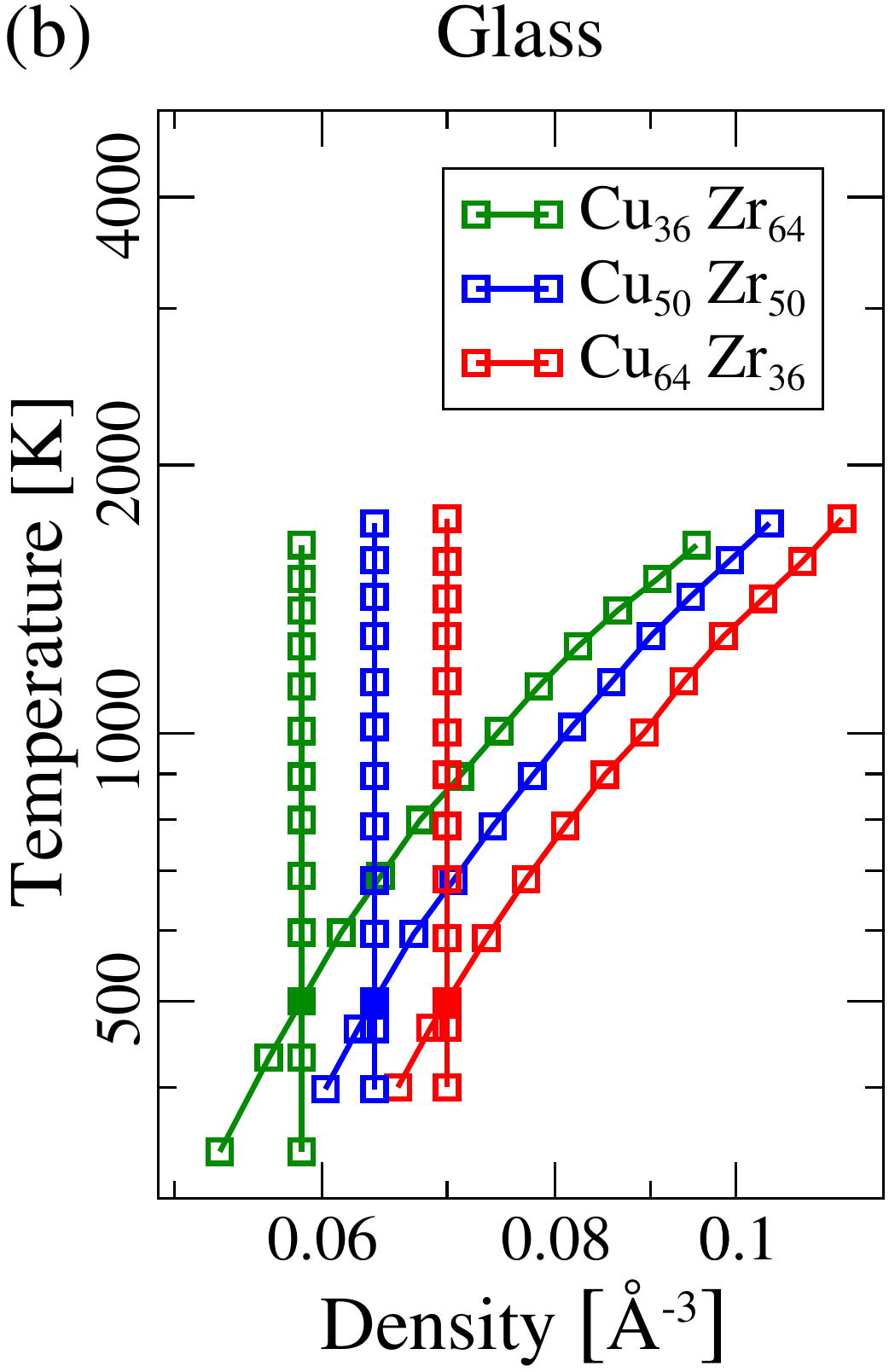}
	\caption{Logarithmic density-temperature phase diagrams showing all state points simulated along isochores (vertical lines) and isomorphs (lines at an angle). The colors reflect the three different compositions studied. (a) gives liquid state points, (b) gives glass state points. Each isomorph is generated by means of the direct-isomorph-check method (see the text), proceeding in steps of 5\% density changes starting from ``reference'' state points at temperature 1500 K for the liquids and 500 K for the glasses (full symbols). The reference state point densities were selected to have approximately the same pressure ($\sim 17$GPa), while the lowest-density state points have approximately zero pressure (compare Tables \ref{table2} and \ref{table3}). The isochores studied below for comparison to the isomorphs are the vertical lines through each reference state point.}	\label{fig1}
\end{figure}

\subsection{Tracing out isomorphic state points}\label{tracing}

Three compositions were studied, Cu$_{36}$Zr$_{64}$, Cu$_{50}$Zr$_{50}$, and Cu$_{64}$Zr$_{36}$. \Fig{fig1} presents the state points simulated in a density-temperature thermodynamic phase diagram. Isomorph invariance is never perfect in realistic systems. In order to estimate to which degree this  invariance holds, it is therefore useful to compare structure and dynamics variations along isomorphs to what happens along curves of similar temperature or density variation, which are not isomorphs. We have chosen to compare to isochores (lines of constant density) with the same temperature variation as the isomorphs. The isochores, of course, are the vertical straight lines in \fig{fig1}, the isomorphs are the lines with a slope. The high-temperature state points describe equilibrium liquids, the low-temperature points are glass-phase state points.

\begin{figure}[t]
	\centering
	\includegraphics[width=6cm]{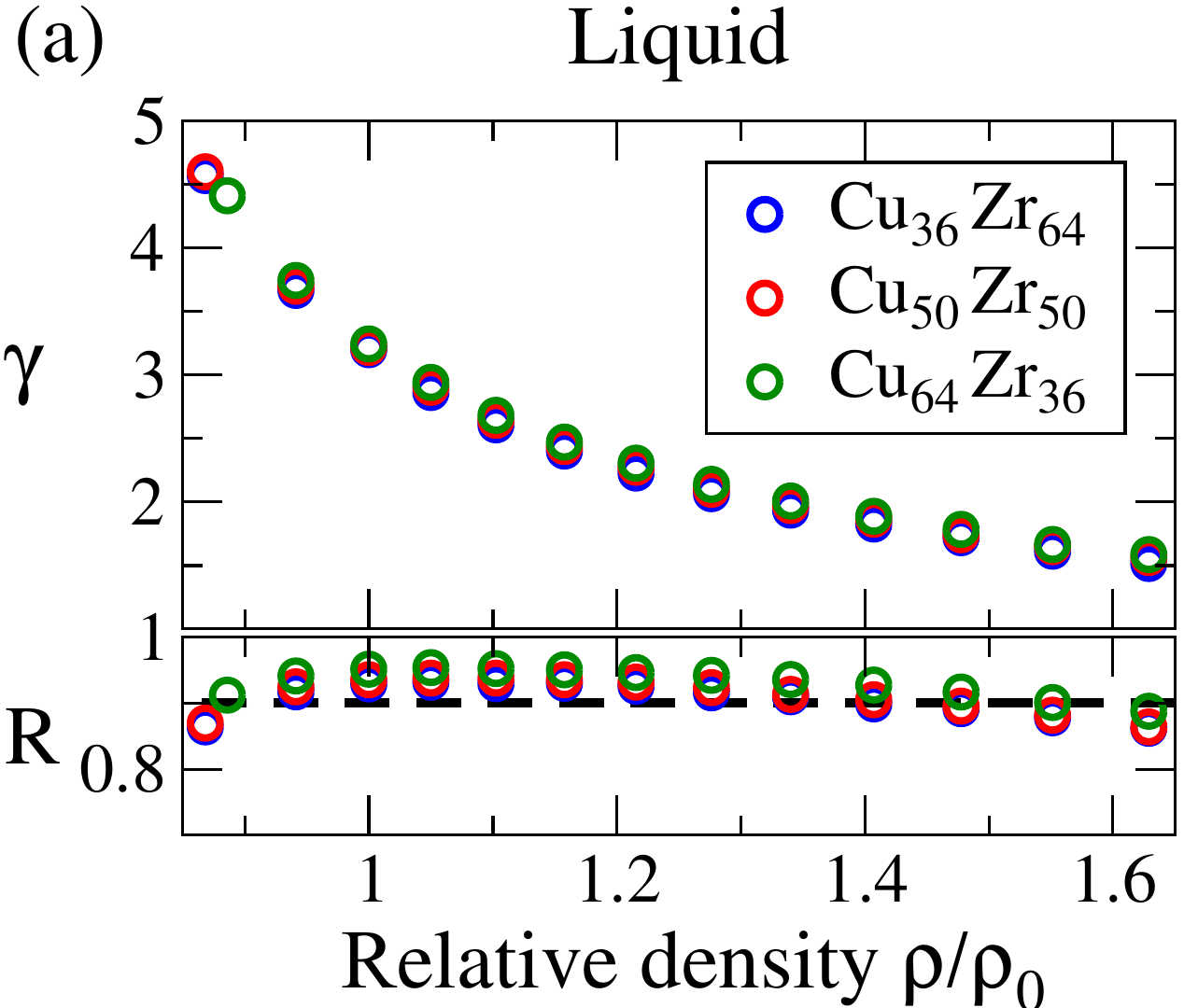}
	\includegraphics[width=6cm]{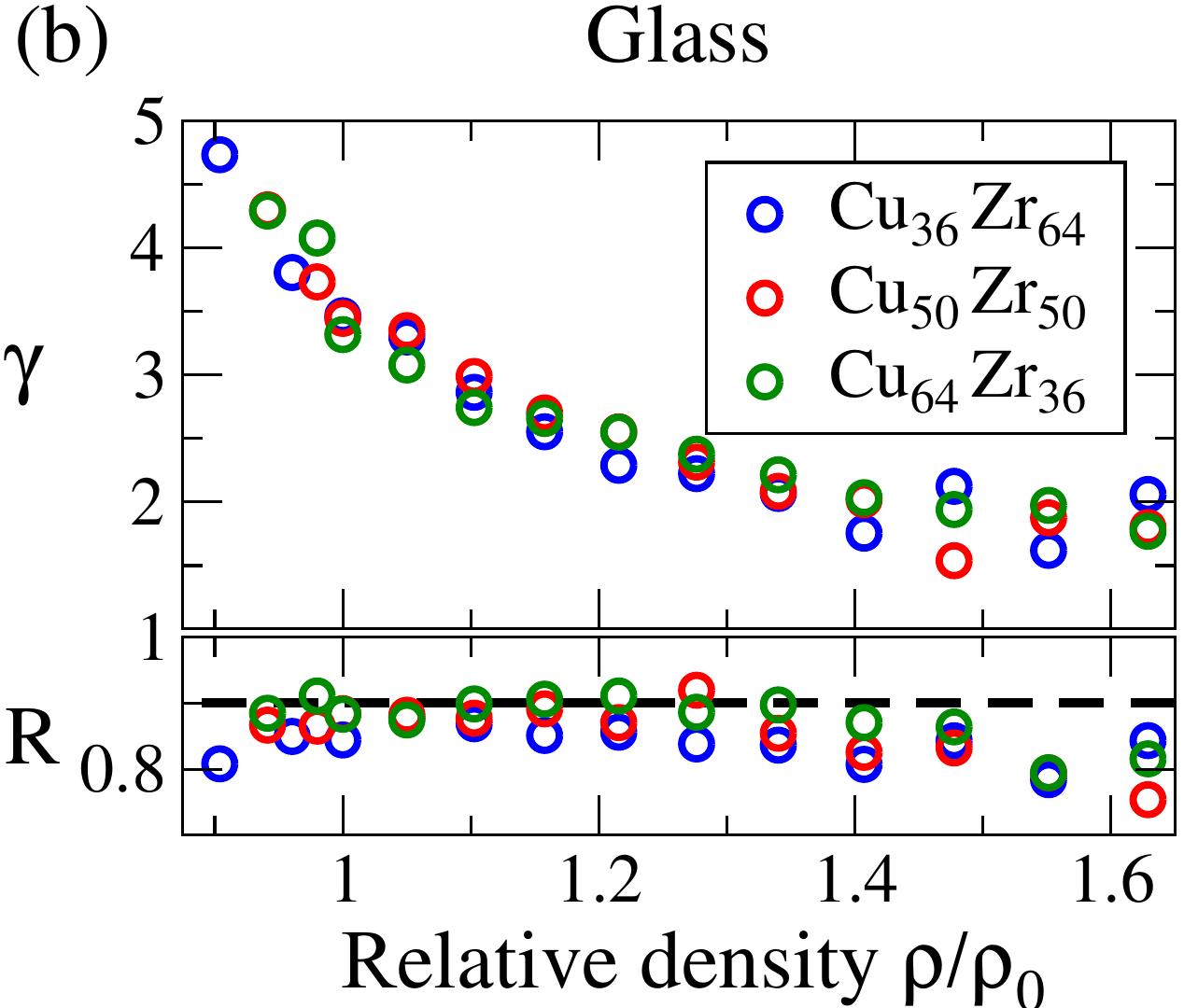}
	\caption{Virial potential-energy Pearson correlation coefficient $R$ (\eq{R_def}) and density-scaling exponent $\gamma$ (\eq{gamma}), plotted along the isomorphs as a function of the density relative to that of the reference state point (which has temperature 1500 K for liquids and 500 K for glasses). (a) is the liquid, (b) is the glass. We see similar pictures in the two cases, with $\gamma$ decreasing significantly as density is increased, indicating an effective softening of the interactions. The virial potential-energy correlations are generally strong, with a maximum at densities close to the reference state point densities denoted by $\rho_0$. The dashed lines mark $R=0.9$, which is traditionally used for delimiting state points for which isomorph-theory predictions are expected to apply \cite{I,IV}.}
	\label{fig2}
\end{figure}

We now turn to the challenge of tracing out isomorphs. Recall that an isomorph is a curve of constant $\Sex$ for a system that obeys the hidden-scale-invariance condition \eq{hsi} at the relevant state points. To which degree this condition is obeyed may be difficult to judge because \eq{hsi} always applies when $\lambda$ is close to unity, but fortunately a practical criterion exists: \eq{hsi} applies to a good approximation if and only if the virial $W$ and potential energy $U$ are strongly correlated in their thermal-equilibrium constant-density ($NVT$) fluctuations \cite{sch14}. These fluctuations are characterized by the Pearson correlation coefficient $R$ defined by (in which sharp brackets denote canonical-ensemble averages and $\Delta$ is the deviation from the thermal average)

\be\label{R_def}
R
\,=\,\frac{\langle\Delta U\Delta W\rangle}{\sqrt{\langle(\Delta U)^2\rangle\langle(\Delta W)^2\rangle}}\,.
\ee
As a pragmatic criterion, $R>0.9$ is usually used for delimiting where isomorph-theory predictions are expected to apply \cite{I,IV,dyr14}. For the CuZr systems we find that $R$ goes below 0.9 at high densities in the liquid phase, as well as in most of the glass phase (\fig{fig2}), but at most state points studied $R$ is above 0.8. Thus it makes good sense to test for isomorph invariance.

\begin{figure}[htbt!]
	\centering
	\includegraphics[width=6cm]{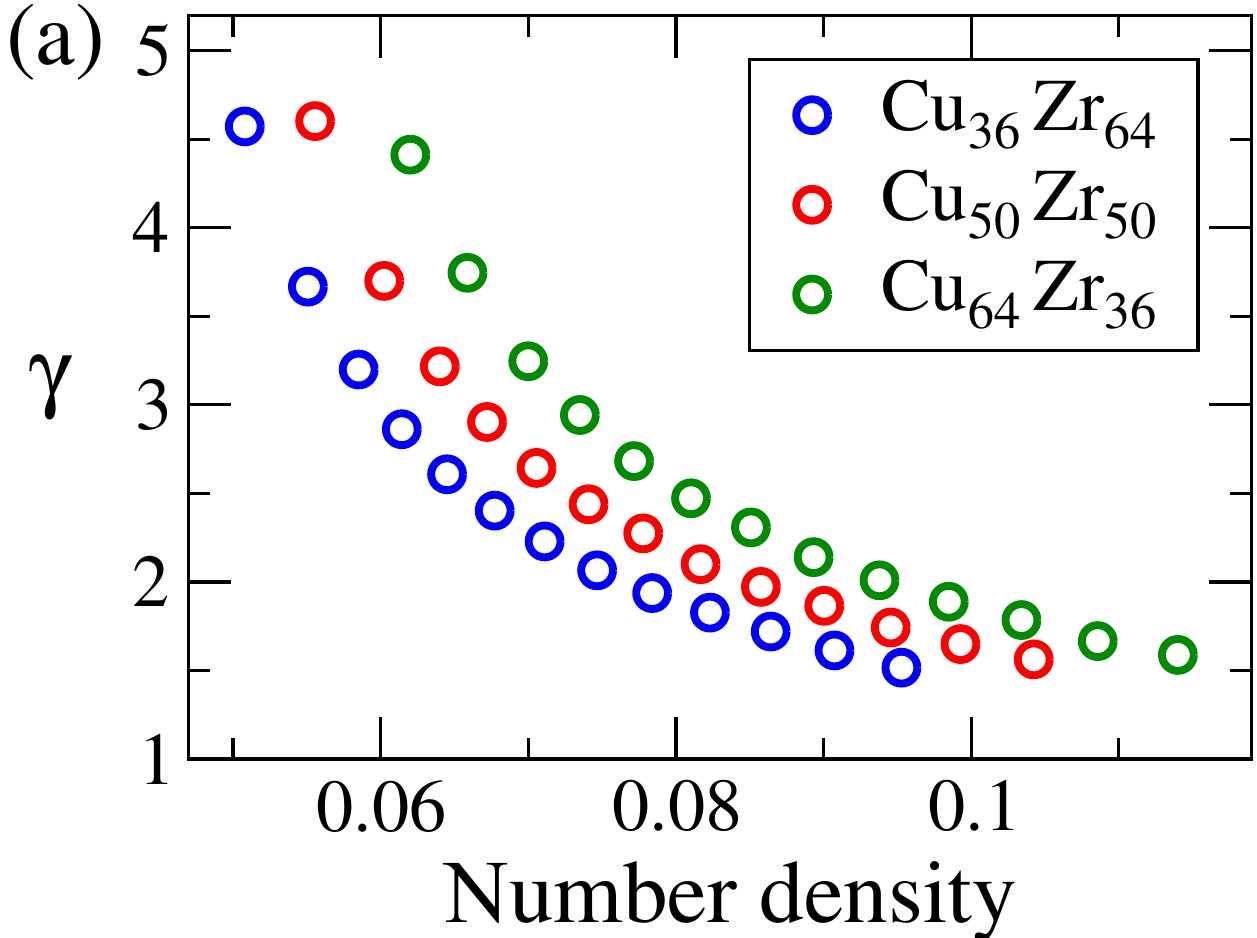}
	\includegraphics[width=6cm]{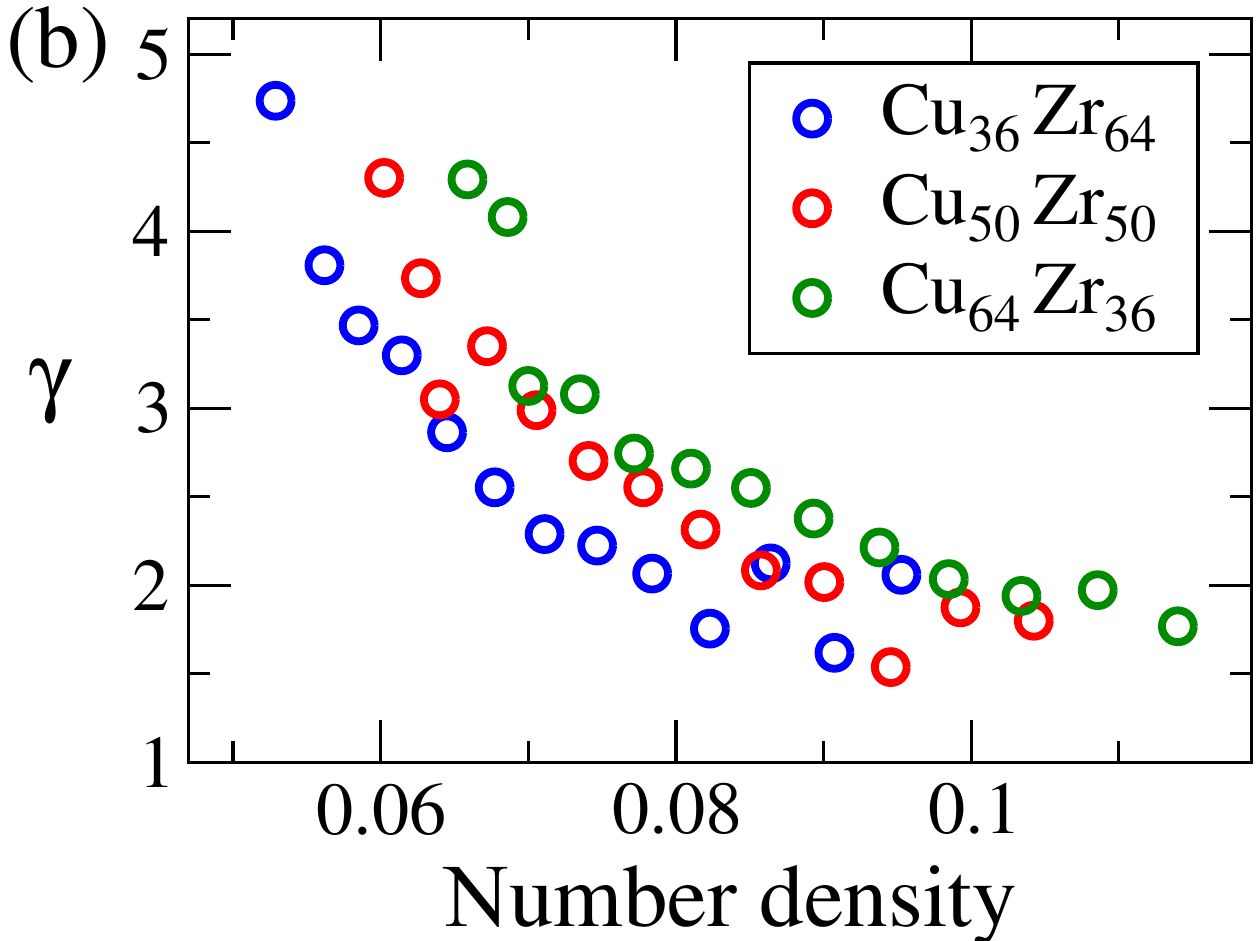}
	\includegraphics[width=6cm]{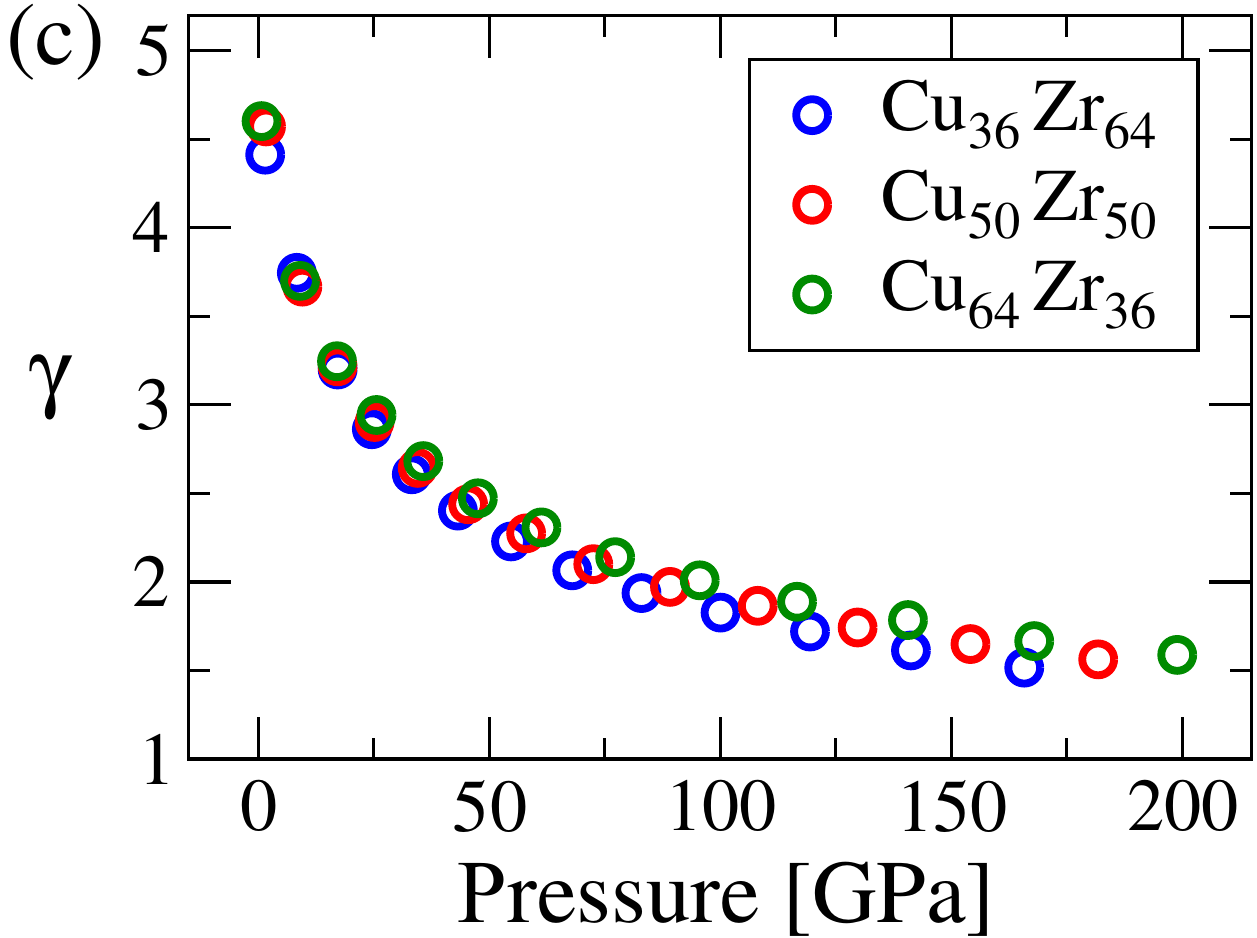}
	\includegraphics[width=6cm]{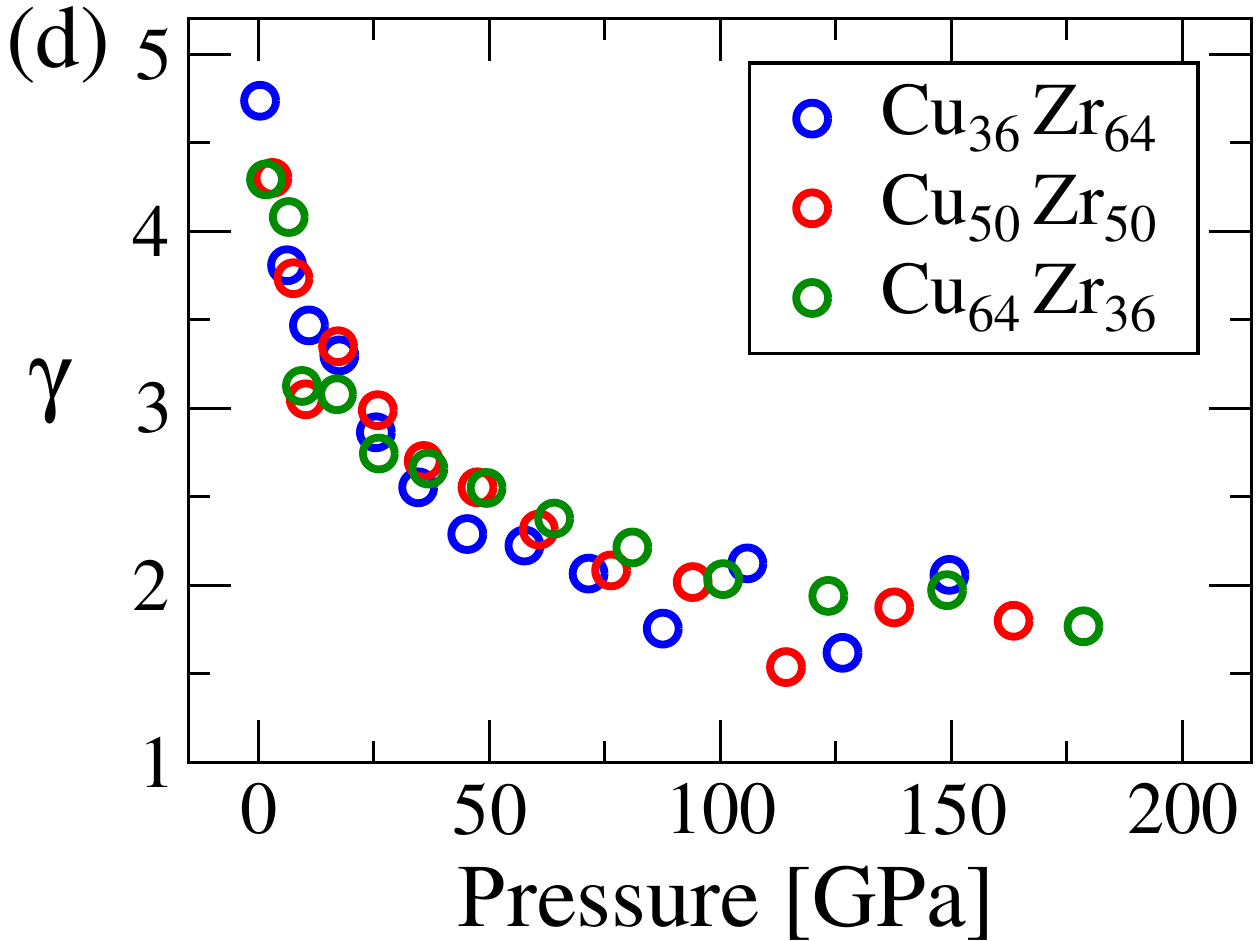}
	\caption{The density-scaling exponent $\gamma$ for all state point studied plotted as a function of density and pressure. (a) and (b) show data for $\gamma$ as a function of the density for the liquid and glass state points, respectively, (c) and (d) similarly show $\gamma$ as a function of the pressure.}
	\label{fig2a}
\end{figure}

Tracing out a curve of constants excess entropy is straightforward if one knows how $\Sex$ varies throughout the phase diagram. It is a bit challenging to evaluate entropy, however, because this involves thermodynamic integration (or the Widom insertion method that is also tedious). In order to trace out an isomorph, one does not need to know the value of $\Sex$, however, and one can therefore make use of the following general identity

\be\label{gamma}
\gamma
\,\equiv\, \left(\frac{\partial\ln T}{\partial\ln\rho}\right)_{\Sex}
\,=\,\frac{\langle\Delta U \Delta W \rangle}{\langle(\Delta U)^2\rangle}\,.
\ee
Here the quantity $\gamma$ is the (state-point dependent) density-scaling exponent defined as the isomorph slope in a logarithmic density-temperature phase diagram like that of \fig{fig1}. The second equality sign is a statistical-mechanical identity that allows for calculating $\gamma$ from $NVT$ equilibrium fluctuations at the state point in question \cite{IV}. \Fig{fig2} shows how $\gamma$ varies along the isomorphs of the CuZr systems studied below, plotted as a function of the density relative to that of the isomorph reference state point. All cases show similar behavior with $\gamma$ decreasing significantly with increasing density. This indicates a softening of the interactions at high densities.

\Fig{fig2a} looks more closely into what controls the density-scaling exponent $\gamma$. The two upper figures show $\gamma$ as a function of the (number) density for, respectively, the liquid and glass state points. While there was a good collapse when $\gamma$ was plotted as a function of density relative to the reference-state-point density (\fig{fig2}), this is no more the case. However, plotting $\gamma$ as a function of the pressure results in an approximate data collapse (lower figures). Interestingly, this finding is consistent with a recent conjecture by Casalini and Ransom that was formulated in the entirely different context of supercooled organic liquids \cite{cas20}. The glass data are more noisy than the liquid data, which we ascribe to the fact that a glass consists of atoms vibrating in a single potential-energy minimum, i.e., just a single so-called inherent state is monitored.

\Eq{gamma} can be used to trace out an isomorph by numerical integration, for instance using the Euler algorithm for small density changes of order one percent \cite{IV} or using the fourth-order Runge-Kutta algorithm that allows for significantly larger density changes \cite{att21}. While both methods are accurate, they both involve many simulations if one wishes to cover a significant density range. Fortunately, there are alternative computationally more efficient methods. For instance, isomorphs of the Lennard-Jones system are to a good approximation given by $h(\rho)/T=$Const. in which $h(\rho)=(\gamma_0/2-1)(\rho/\rho_0)^4-(\gamma_0/2-2)(\rho/\rho_0)^2$ in which $\gamma_0$ is the density-scaling exponent at a selected ``reference state point'' of density $\rho_0$ \cite{boh12,ing12a}.

\begin{figure}[t]
	\centering
	\includegraphics[width=7cm]{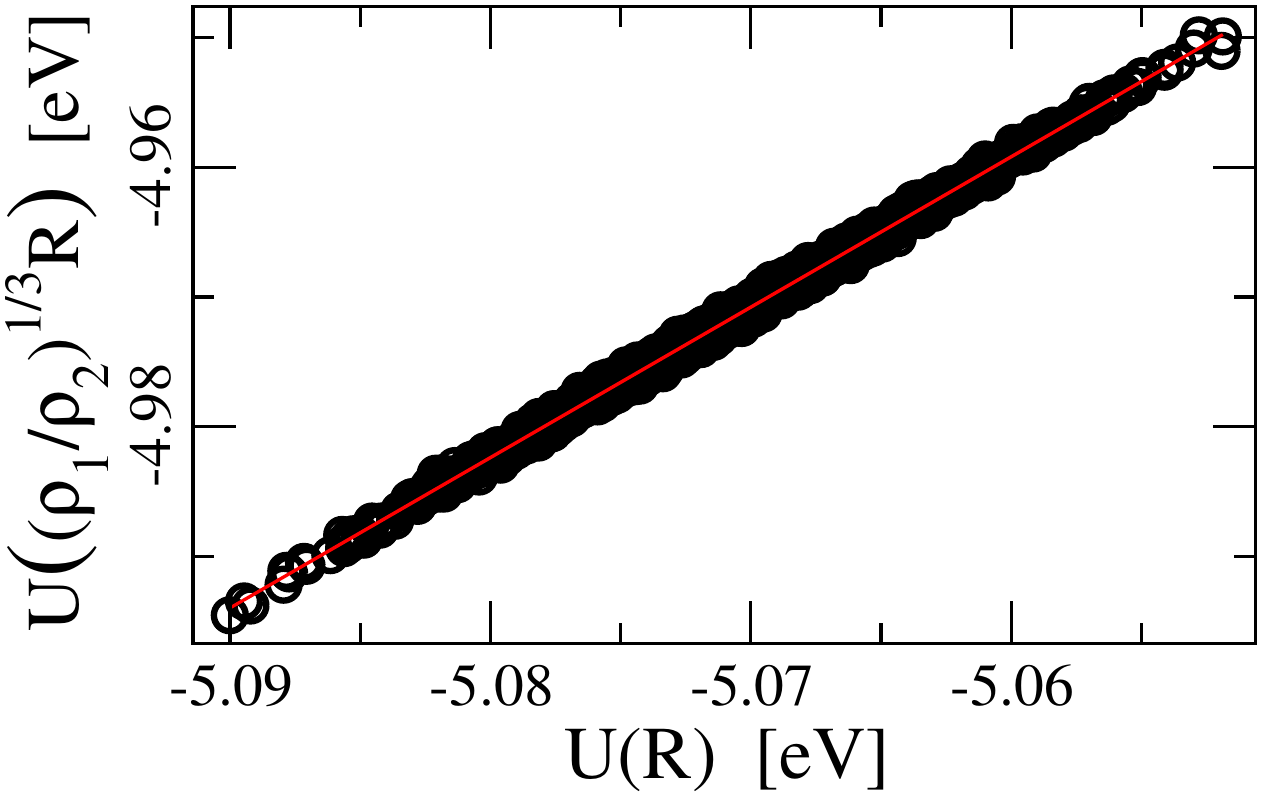}
	\caption{Example of a direct isomorph check (DIC), here with state point $1$ being the reference state point of the 64-36 mixture ($T=1500$K) and state point $2$ having a 5\% higher density. At state point 1, a series of thermal-equilibrium configurations $\bR$ are sampled. Each of these are scaled uniformly by the factor $(\rho_2/\rho_1)^{-1/3}=0.9839$, resulting in a scaling to a 5\% higher density. Plotting the potential energy of scaled versus unscaled configuration, the slope of the best fit line is $T_2/T_1$, compare \eq{prop_fluct}; this determines the temperature $T_2$ that makes the state point $(\rho_2,T_2)$ isomorphic to state point $(\rho_1,T_1)$.}
	\label{fig3}
\end{figure}

\begin{table*}[t]
	\centering
	\begin{tabular}{r r r | r r r | r r r}
		$T$ [K] & $\rho$ [\AA${}^{-3}$] & $p$ [GPa] & $T$ [K] & $\rho$ [\AA${}^{-3}$] & $P$ [GPa] & $T$ [K] & $\rho$ [\AA${}^{-3}$] & $P$ [GPa]  \\ \hline \hline
		875 &  0.0508 &   1.7 &  870 &  0.0556 &   0.8 &  950 &  0.0620 &   1.5 \\
		1219 &  0.0551 &   9.6 & 1217 &  0.0602 &   9.0 & 1215 &  0.0659 &   8.4 \\
		\textbf{1500} &  \textbf{0.0585} &  \textbf{17.2} & \textbf{1500} &  \textbf{0.0640} &  \textbf{17.2} & \textbf{1500} &  \textbf{0.0700} &  \textbf{17.1} \\
		 1745 &  0.0614 &  24.6 & 1747 &  0.0672 &  25.2 & 1749 &  0.0735 &  25.6 \\
		1999 &  0.0645 &  33.2 & 2005 &  0.0706 &  34.5 & 2009 &  0.0772 &  35.7 \\
		2266 &  0.0677 &  43.2 & 2274 &  0.0741 &  45.4 & 2283 &  0.0810 &  47.6 \\
		2540 &  0.0711 &  54.8 & 2555 &  0.0778 &  58.0 & 2568 &  0.0851 &  61.3 \\
		2828 &  0.0747 &  68.0 & 2850 &  0.0817 &  72.5 & 2867 &  0.0893 &  77.3 \\
		3122 &  0.0784 &  83.0 & 3151 &  0.0858 &  89.2 & 3177 &  0.0938 &  95.6 \\
		3428 &  0.0823 & 100.1 & 3464 &  0.0901 & 108.1 & 3499 &  0.0985 & 116.6 \\
		3746 &  0.0864 & 119.5 & 3795 &  0.0946 & 129.7 & 3834 &  0.1034 & 140.7 \\
		4070 &  0.0908 & 141.3 & 4128 &  0.0993 & 154.2 & 4179 &  0.1086 & 168.0 \\
		4404 &  0.0953 & 165.8 & 4473 &  0.1042 & 181.8 & 4533 &  0.1140 & 199.0 \\

	\end{tabular}
	\caption{Temperature, density, and pressure of liquid isomorph state points.The vertical lines divide the compositions in the following order: Cu$_{36}$Zr$_{64}$, Cu$_{50}$Zr$_{50}$, Cu$_{64}$Zr$_{36}$. The boldface row represents the reference state point for the isomorph of each composition. }
	\label{table2}
\end{table*}

\begin{table*}[t]
	\centering
	\begin{tabular}{r r r | r r r | r r r}
$T$ [K] & $\rho$ [\AA${}^{-3}$] & $p$ [GPa] & $T$ [K] & $\rho$ [\AA${}^{-3}$] & $P$ [GPa] & $T$ [K] & $\rho$ [\AA${}^{-3}$] & $P$ [GPa]  \\ \hline \hline
 339 &  0.0529 &   0.4 &  398 &  0.0602 &   3.1 &  400 &  0.0659 &   1.7 \\
432 &  0.0562 &   6.2 &  466 &  0.0627 &   7.7 &  467 &  0.0686 &   6.6 \\
\textbf{500} &  \textbf{0.0585} &  \textbf{11.0} &  \textbf{500} &  \textbf{0.0640} &  \textbf{10.3} & \textbf{500} &  \textbf{0.0700} &   \textbf{9.5} \\
587 &  0.0614 &  17.6 &  575 &  0.0672 &  17.3 & 578 &  0.0735 &  17.1 \\
683 &  0.0645 &  25.5 &  672 &  0.0706 &  25.9 & 667 &  0.0772 &  26.1 \\
781 &  0.0677 &  34.6 &  773 &  0.0741 &  35.8 &  758 &  0.0810 &  36.8 \\
880 &  0.0711 &  45.2 &  877 &  0.0778 &  47.5 &  858 &  0.0851 &  49.4 \\
980 &  0.0747 &  57.6 &  989 &  0.0817 &  60.7 &  968 &  0.0893 &  64.1 \\
1088 &  0.0784 &  71.5 &  1103 &  0.0858 &  76.4 &  1083 &  0.0938 &  81.0 \\
1200 &  0.0823 &  87.6 &  1218 &  0.0901 &  94.1 &  1203 &  0.0985 & 100.6 \\
1304 &  0.0864 & 105.9 &  1340 &  0.0946 & 114.3 &  1325 &  0.1034 & 123.4 \\
1442 &  0.0908 & 126.5 &  1440 &  0.0993 & 137.7 &  1453 &  0.1086 & 149.1 \\
1558 &  0.0953 & 149.7 &  1575 &  0.1042 & 163.5 &  1596 &  0.1140 & 178.7 \\  
\end{tabular}
	\caption{Temperature, density, and pressure of glass isomorph state points.  The vertical lines divide the compositions shown in the following order: Cu$_{36}$Zr$_{64}$, Cu$_{50}$Zr$_{50}$, Cu$_{64}$Zr$_{36}$. The boldface row represents the reference state point for the isomorph of each composition. The starting point for the glass isomorph has the same density as that of the liquid isomorph, but temperature $500$K.}
	\label{table3}
\end{table*}

A general and fairly efficient method for tracing out isomorphs is the ``direct isomorph check'' (DIC) \cite{IV}, and we used this for generating the CuZr isomorphs. The DIC is justified as follows \cite{sch14}. Hidden scale invariance (\eq{hsi}) implies that the microscopic excess entropy function $\Sex(\bR)$ is scale invariant, i.e., a function only of a configuration's reduced coordinate $\tbR$: $\Sex(\bR)=\Sex(\tbR)$ \cite{sch14}. From the definition of $\Sex(\bR)$ it follows that $U(\bR)=U(\rho,\Sex(\tbR))$ in which the function $U(\rho,\Sex)$ is the average potential energy at the state point with density $\rho$ and excess entropy $\Sex$ \cite{sch14}. Considering configurations with the same density $\rho$ and small deviations in the microscopic excess entropy from that of the given state point $\Sex$, an expansion to first order leads to

\be\label{Ueq}
U(\bR)
\,\cong\, U(\rho,\Sex) +T(\rho,\Sex) \left(\Sex(\tbR)-\Sex\right)\,.
\ee
Consider two state points $(\rho_1,T_1)$ and $(\rho_2,T_2)$ with the same excess entropy $\Sex$ and average potential energies $U_1$ and $U_2$, respectively. If $\bR_1$ and $\bR_2$ are configurations of these state points with the same reduced coordinates, i.e., obeying

\be
\rho_1^{1/3}\bR_1=\rho_2^{1/3}\bR_2\equiv \tbR\,,
\ee
one gets by elimination of the common factor $\Sex(\tbR)-\Sex$ in \eq{Ueq} with $T_1\equiv T(\rho_1,\Sex)$ and $T_2\equiv T(\rho_2,\Sex)$

\be\label{isom}
\frac{U(\bR_1)-U_1}{T_1}\,\cong\,\frac{U(\bR_2)-U_2}{T_2}\,.
\ee
While not of direct relevance for the present paper, we note that \eq{isom} implies $\exp(-{U(\bR_1)}/{k_B T_1})\propto \exp(-{U(\bR_2)}/{k_B T_2})$, i.e., that the two configurations have the same canonical probability. This is a manifestation of the hidden scale invariance inherent in isomorph theory \cite{IV}. 

\Eq{isom} leads to $U (\bR_2)\cong({T_2}/{T_1})U (\bR_1)+\left(U_2-({T_2}/{T_1})U_1\right)$. For the fluctuations about the respective mean values this implies

\be\label{prop_fluct}
\Delta U(\bR_2)
\,\cong\,\frac{T_2}{T_1}\, \Delta U (\bR_1) \,.
\ee
\Eq{prop_fluct} implies that isomorphic state points may be identified as follows: First sample a set of equilibrium configurations at the state point $(\rho_1, T_1)$. Then scale these configurations uniformly to density $\rho_2$. The temperature $T_2$ of the state point with density $\rho_2$, which is isomorphic to state point $(\rho_1, T_1)$, is now found from the slope of a scatter plot of the potential energies of scaled versus unscaled configurations. An example of how this works is shown in \fig{fig3}.

\begin{figure}[t]
	\includegraphics[width=5cm]{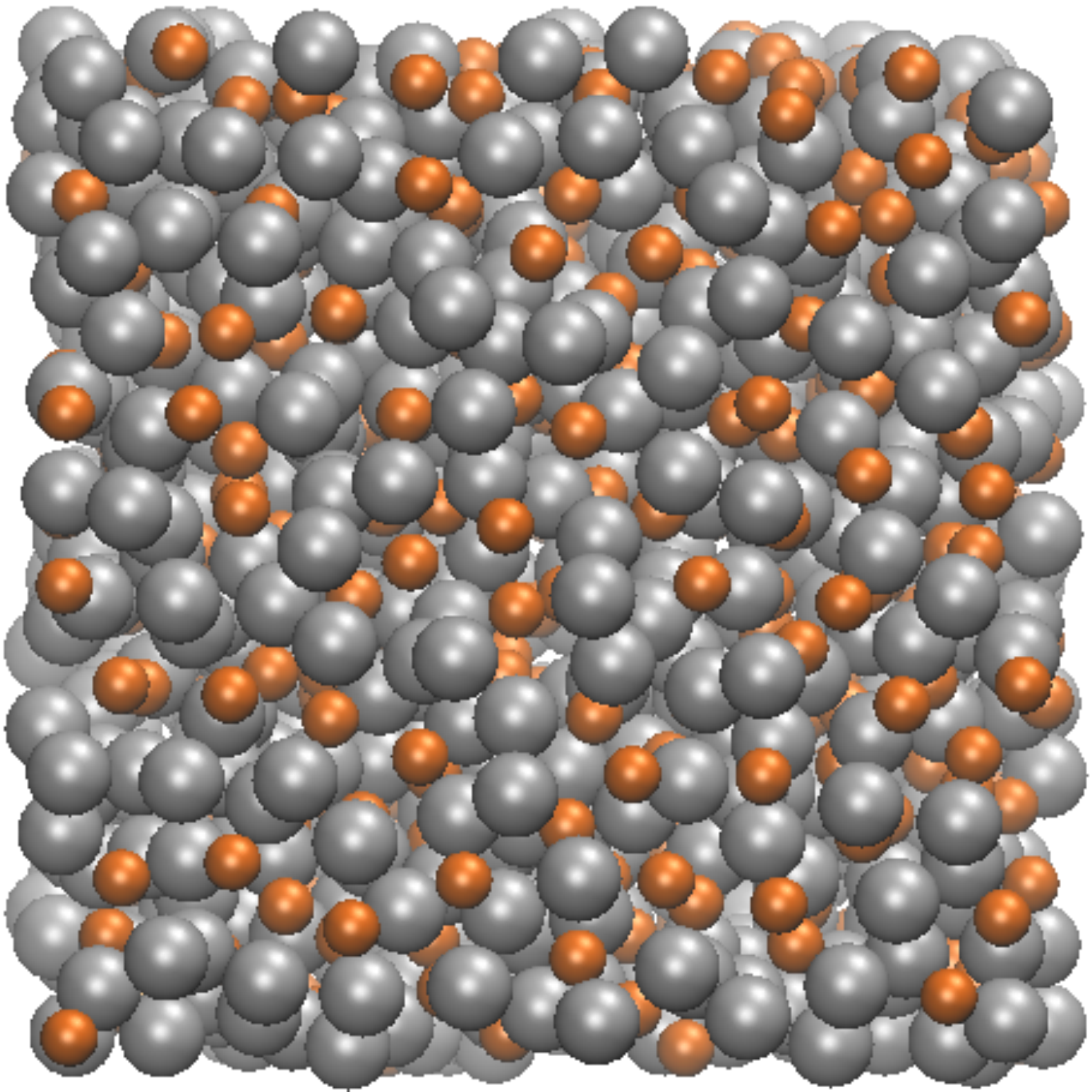}
	\includegraphics[width=5cm]{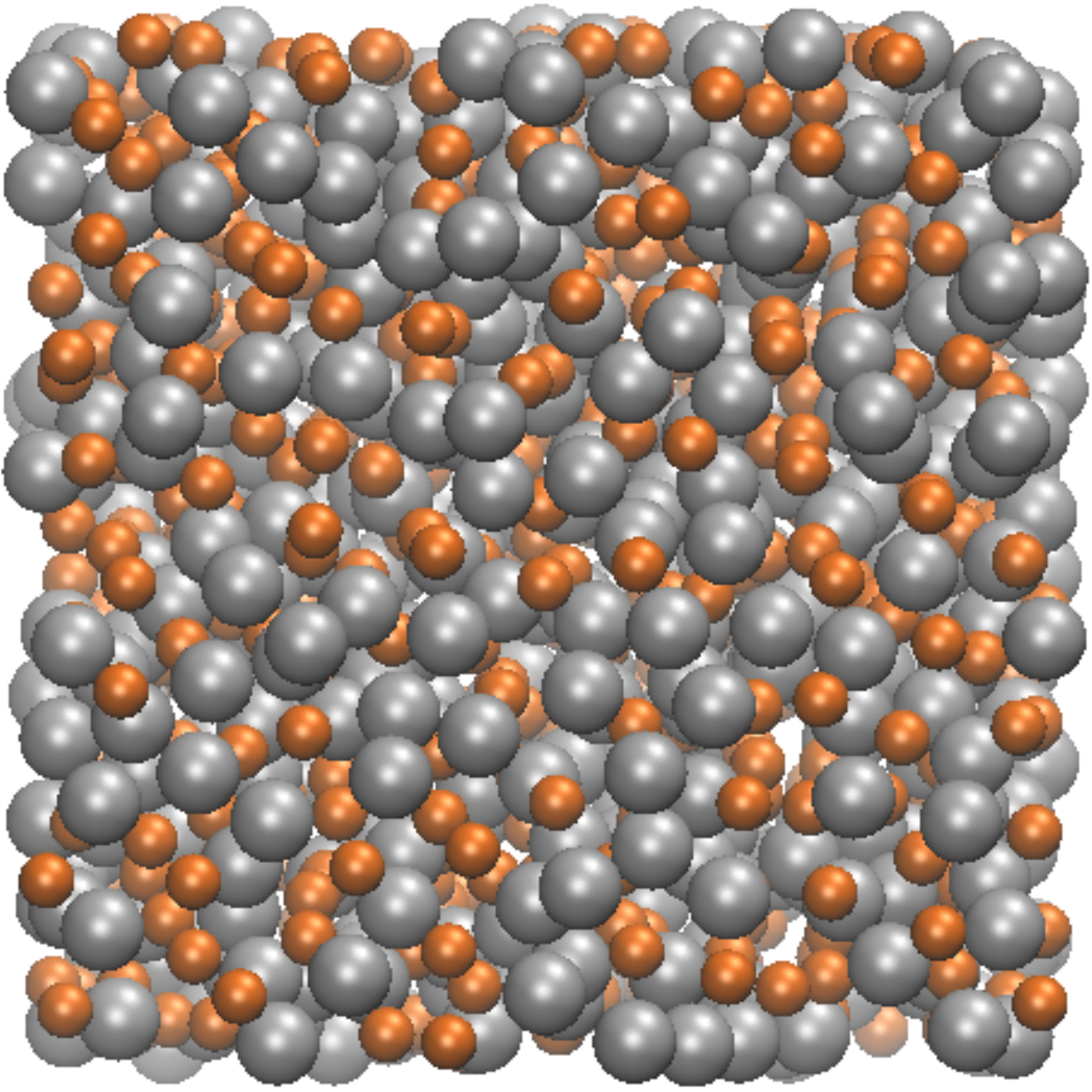}
	\includegraphics[width=5cm]{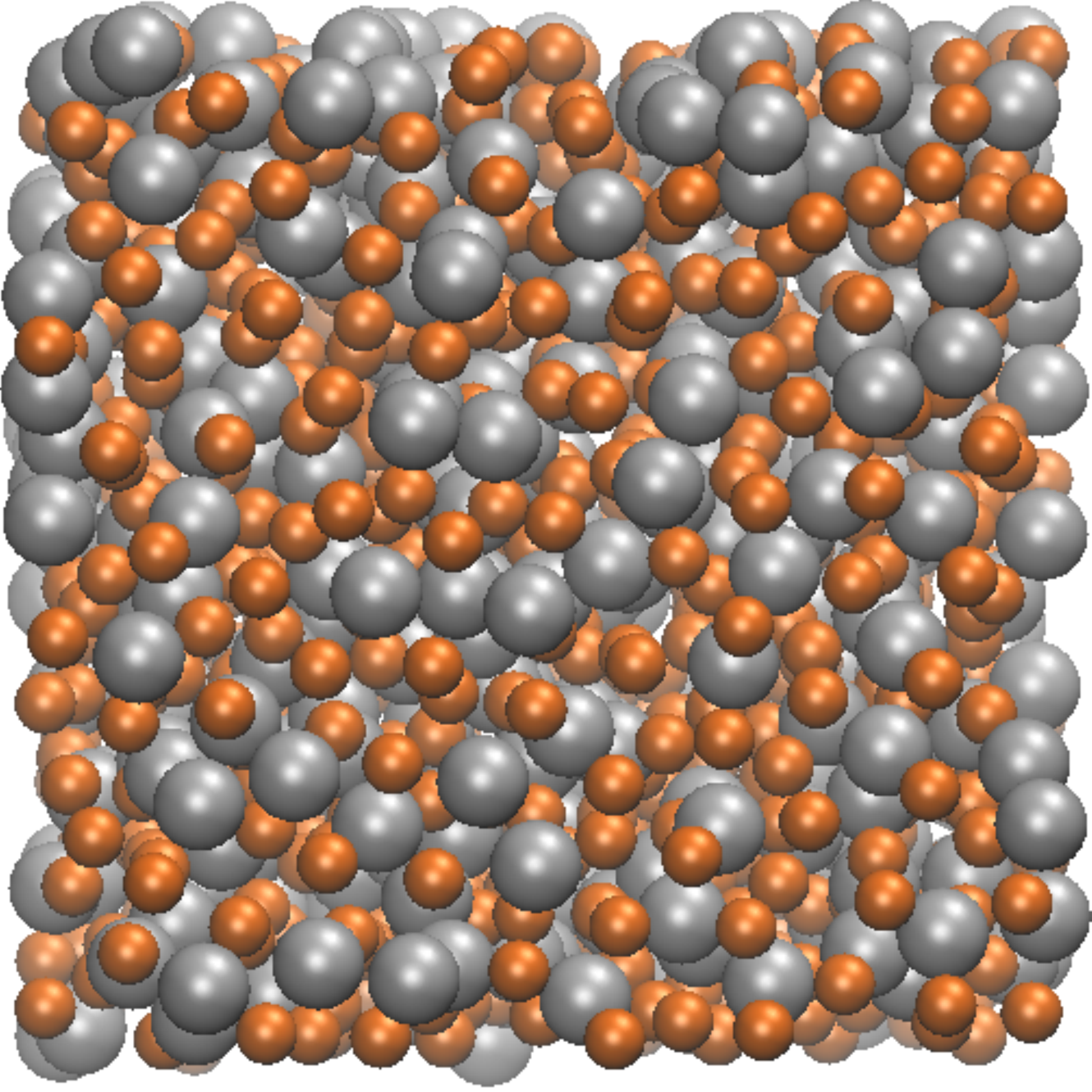}	
	\caption{Snapshots of the glass configurations at the glass isomorph reference state points at 500 K. The Cu atoms are orange, the Zr atoms are grey. From left the figures are for the 36-64, 50-50, and 64-36 CuZr compositions.}	\label{fig_pic}
\end{figure}

Because the hidden-scale-invariance property is not exact, the DIC is less reliable for large density changes than for smaller ones. We traced out the isomorphs studied below using step-by-step DICs involving density changes of 5\%. The resulting isomorphs are shown in \fig{fig1}. The simulated isomorphic state points are listed in Table I (liquid) and Table II (glass).

\begin{figure}[t]
	\centering
	\includegraphics[width=0.25\linewidth]{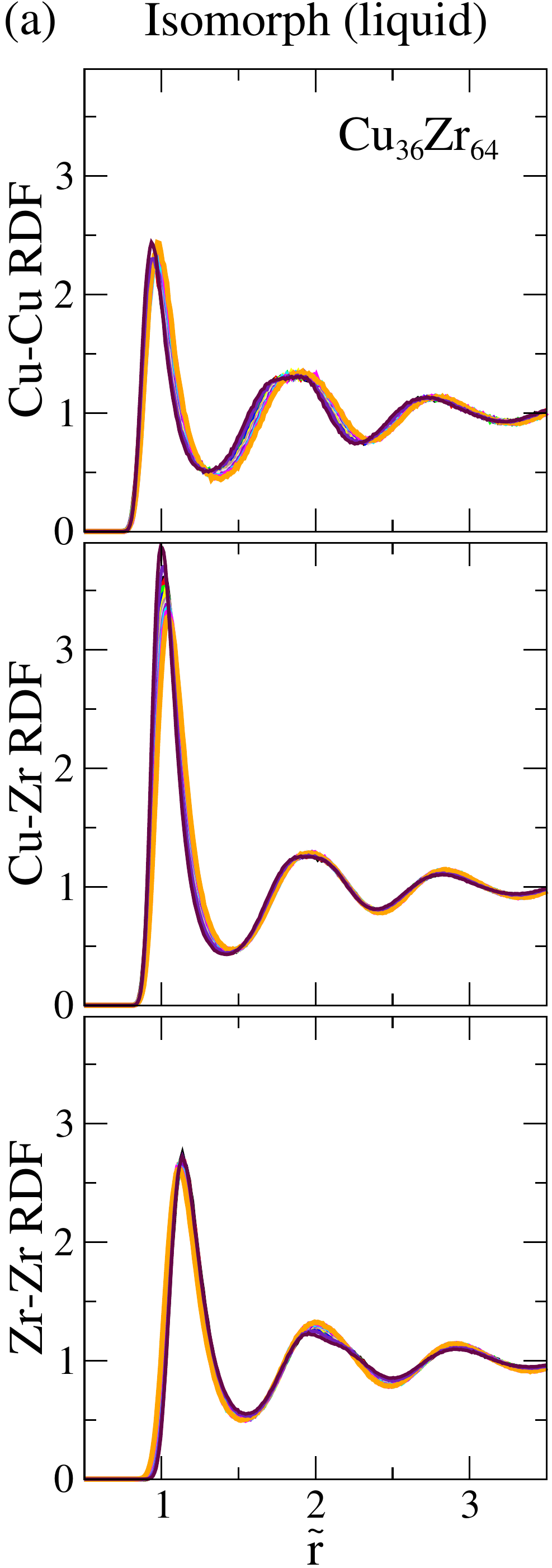}\hfill
	\includegraphics[width=0.25\linewidth]{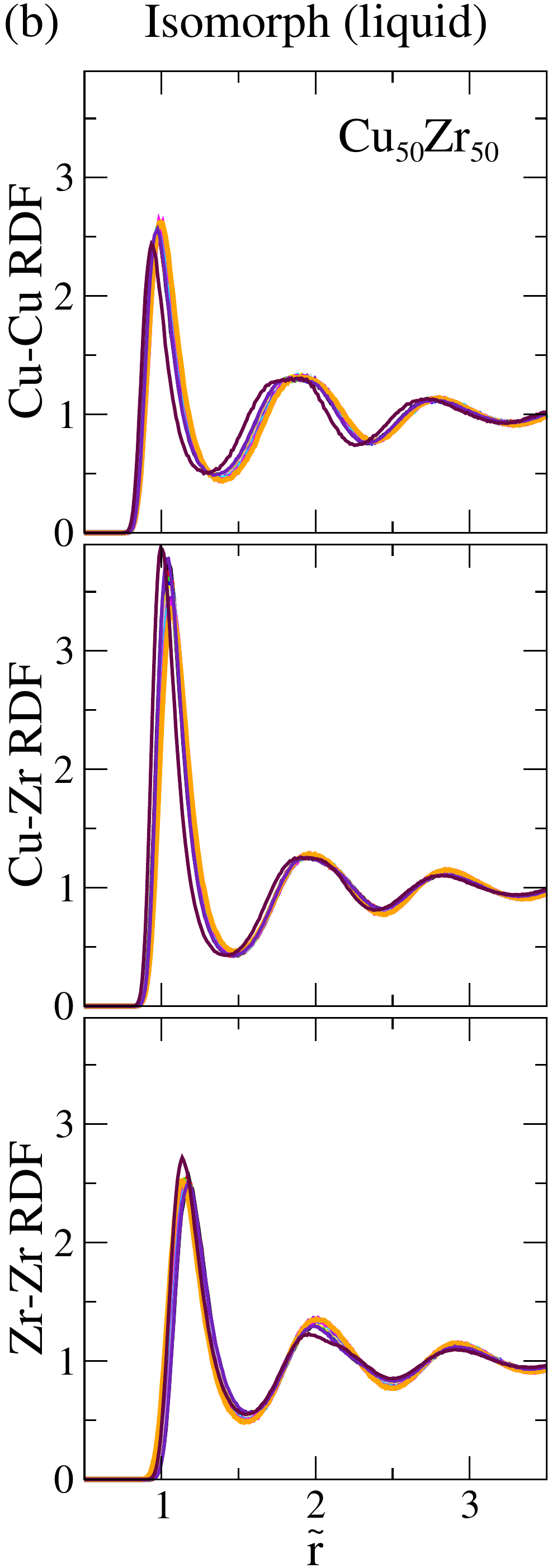}\hfill
	\includegraphics[width=0.25\linewidth]{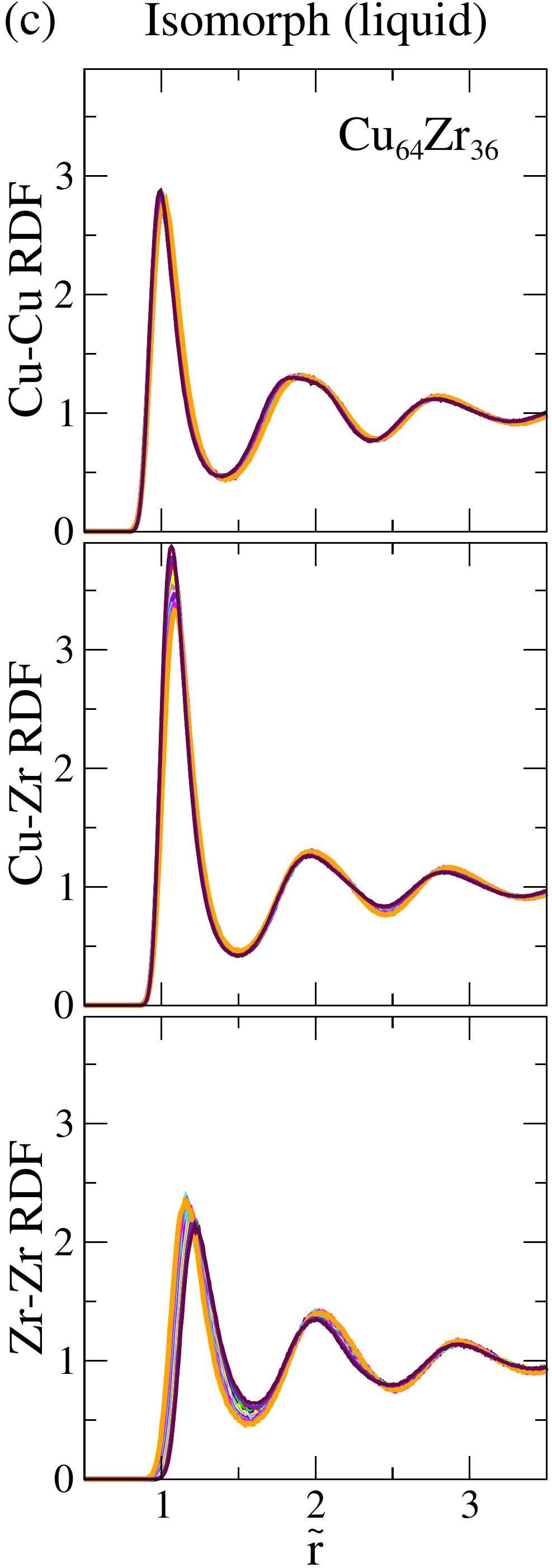}
	\caption{Isomorph liquid-state radial distribution functions (RDFs), plotted as a function of the reduced pair distance $\tilde{r}\equiv\rho^{1/3}r$. (a), (b), and (c) give reduced-unit RDFs along an isomorph for each of the compositions studied. The color coding used is the xmgrace default ordering with black for the data set corresponding to the reference state point, then at increasing density: red, green, blue, yellow, etc; orange is the color given to the 11th data set which is here that the highest temperature. The two state points with lower density than that of the reference state point are purple and brown. Generally, we see good approximate isomorph invariance, with some deviation at the first peak maximum and the largest overall deviations at the lowest densities (at which the virial potential-energy correlation coefficient drops rapidly, compare \fig{fig2}).}
	\label{fig4}
\end{figure}

\section{Simulation details}

The three compositions studied in this work are Cu$_x$Zr$_{100-x}$($x=36,50,64$). For each of these an isomorph was generated from a state point well into the liquid regime. From this initial ``reference'' state point at temperature $1500$ K, an isomorph was traced out using the DIC as described above. The majority of state points are at a higher density than that of the reference state point, $\rho_0$, but for each isomorph we also generated two isomorph state points at lower densities to ensure that samples close to zero pressure were included in the study (compare Tables I and II).

The $NVT$ ensemble implemented via the standard Nos{\'e}-Hoover thermostat was used to simulate cubic boxes containing 1000 particles. For each state point on an isomorph, a state point was simulated at the same temperature at the reference-state-point density; these constitute the isochoric state points discussed below along with the isomorph state points. For all three compositions, glass-phase reference state points were obtained by cooling at a constant rate in 100000 time steps from the liquid reference state point at 1500 K to the glass isomorph reference temperature 500 K. The cooling was implemented by adjusting the Nos{\'e}-Hoover temperature in each step. Since a time step corresponds to 5.1 femtoseconds, this is a very high cooling rate that corresponding roughly to 2 K/picosecond. From these three glass reference-state points, isomorphs were generated by the DIC method in the same way as for the liquids.

The simulations were carried out in RUMD \cite{RUMD}, Roskilde university's GPU Molecular Dynamics package that is optimized for small systems. At each state point the initial configuration was a simple cubic crystal with particle types assigned randomly at the required ratios. At each state point of the liquid, $10^4$ MD steps of equilibration were performed to melt and equilibrate the liquid. Following this we carried out $10^6$ MD steps of the production run. For the glass-phase simulations, $10^5$ MD steps of equilibration were performed before $10^7$ MD steps of the production run. The time step in the simulations was 0.5 $\rm{\AA^{1/2}(u/eV)^{1/2}}$ in which u is the atomic mass unit.

\Fig{fig_pic} shows the glasses prepared by cooling the liquids to the glass-isomorph reference state points. There are no signs of crystallization.

\begin{figure}[t]
	\centering
	\includegraphics[width=0.25\linewidth]{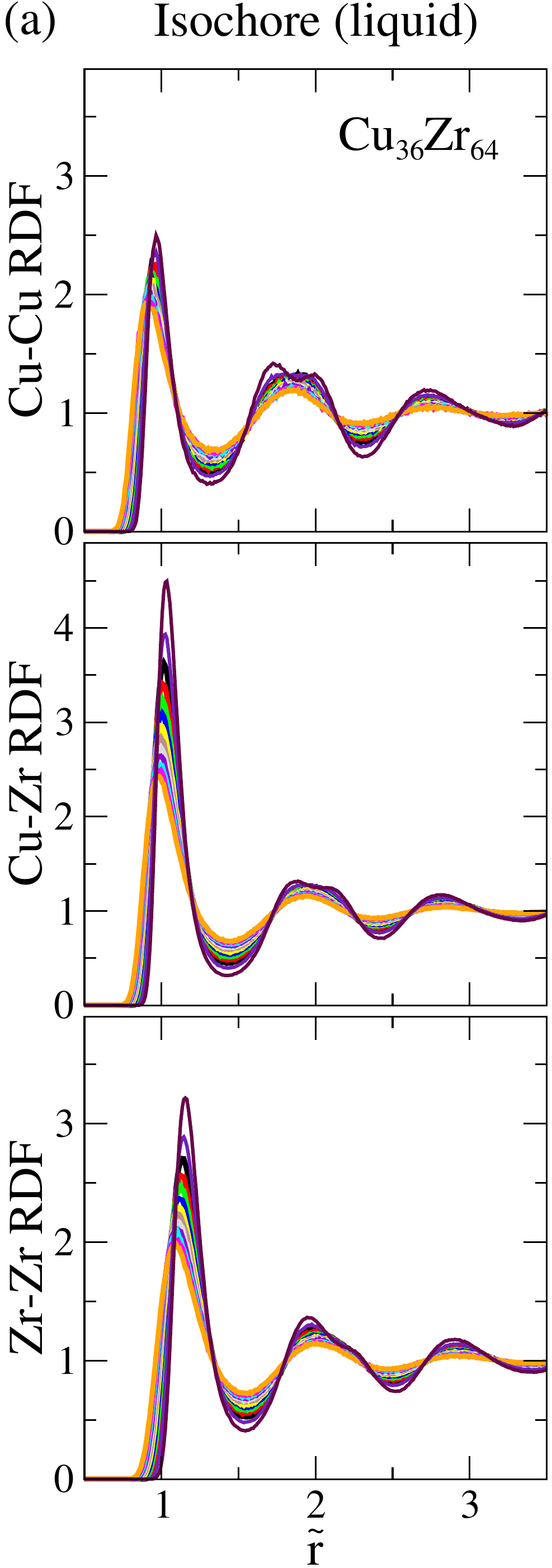}\hfill
	\includegraphics[width=0.25\linewidth]{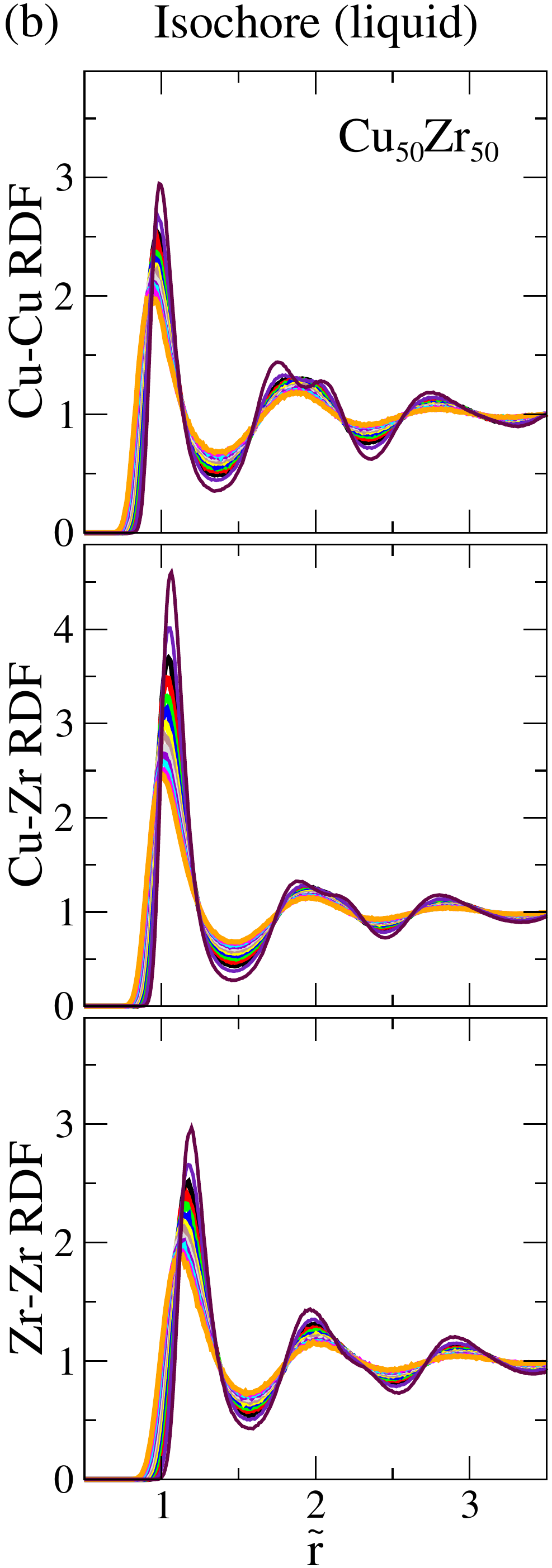}\hfill
	\includegraphics[width=0.25\linewidth]{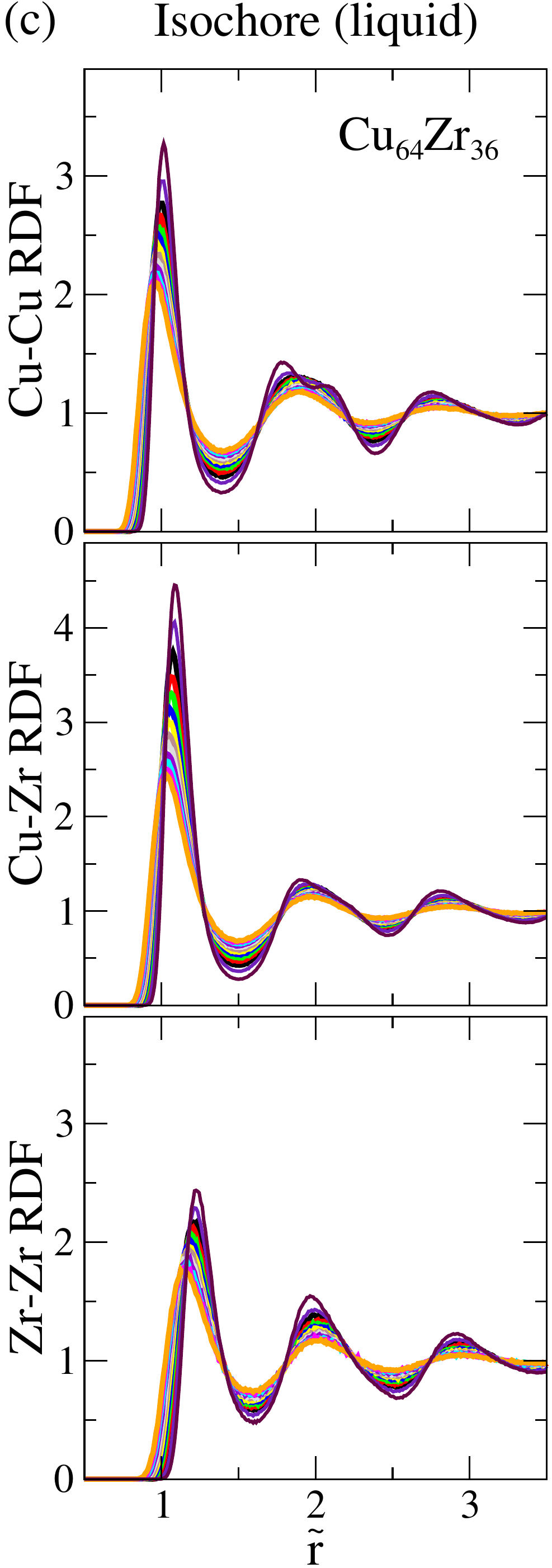}
	\caption{Isochore liquid-state RDFs plotted as in \fig{fig4}. There is a considerably larger variation of the structure, with the structure being more washed out, the higher the temperature is. This reflects the increasing thermal fluctuations, which in the case of an isomorph are compensated by increasing forces (\fig{fig4}).}
	\label{fig5}
\end{figure}

\section{Structure and dynamics in the liquid phase}\label{liquid}

To investigate how the structure varies along isomorphs and isochores for the three CuZr compositions we probed the radial distribution functions (RDFs), which in reduced units are predicted to be isomorph invariant. There are three different RDFs, one for Cu-Cu, one for Cu-Zr, and one for Zr-Zr. Plotting a RDF in reduced units implies scaling the distance variable according to the density (compare \eq{tbReq}). This results in peaks at roughly the same places for all compositions because the scaling corresponds to taking the system to unit density.

\Fig{fig4} shows the reduced RDFs along the isomorphs and \fig{fig5} shows the similar RDFs along isochores with same temperature variation (compare \fig{fig1} showing the similar state points). Comparing the two, we conclude that the structure is isomorph invariant to a good approximation, but varies significantly along the isochores. Deviations are largest for the minority-minority RDFs for the non-equimolar compositions. Deviations from isomorph invariance are also seen in some cases at the first maximum, where the maximum is generally lowered somewhat with increasing  density. This is an effect that is well understood for pair-particle systems, for which it derives from the fact that a higher density-scaling exponent $\gamma$ implies a steeper effective pair potential and therefore less likely particle close encounters. This results in moving some of the low distance RDF to higher distances when $\gamma$ is large, which is the case at low densities. This explanation suggests that $\gamma$ of the present non-pair-potential simulations can also be interpreted as an effective pair IPL exponent \cite{hum15}. -- For the isochores, there is a general ``damping'' of the RDFs at all distances as temperature increases. This reflects the stronger thermal fluctuations at high temperatures. 

\begin{figure*}[t]
	\centering
	\includegraphics[width=8cm]{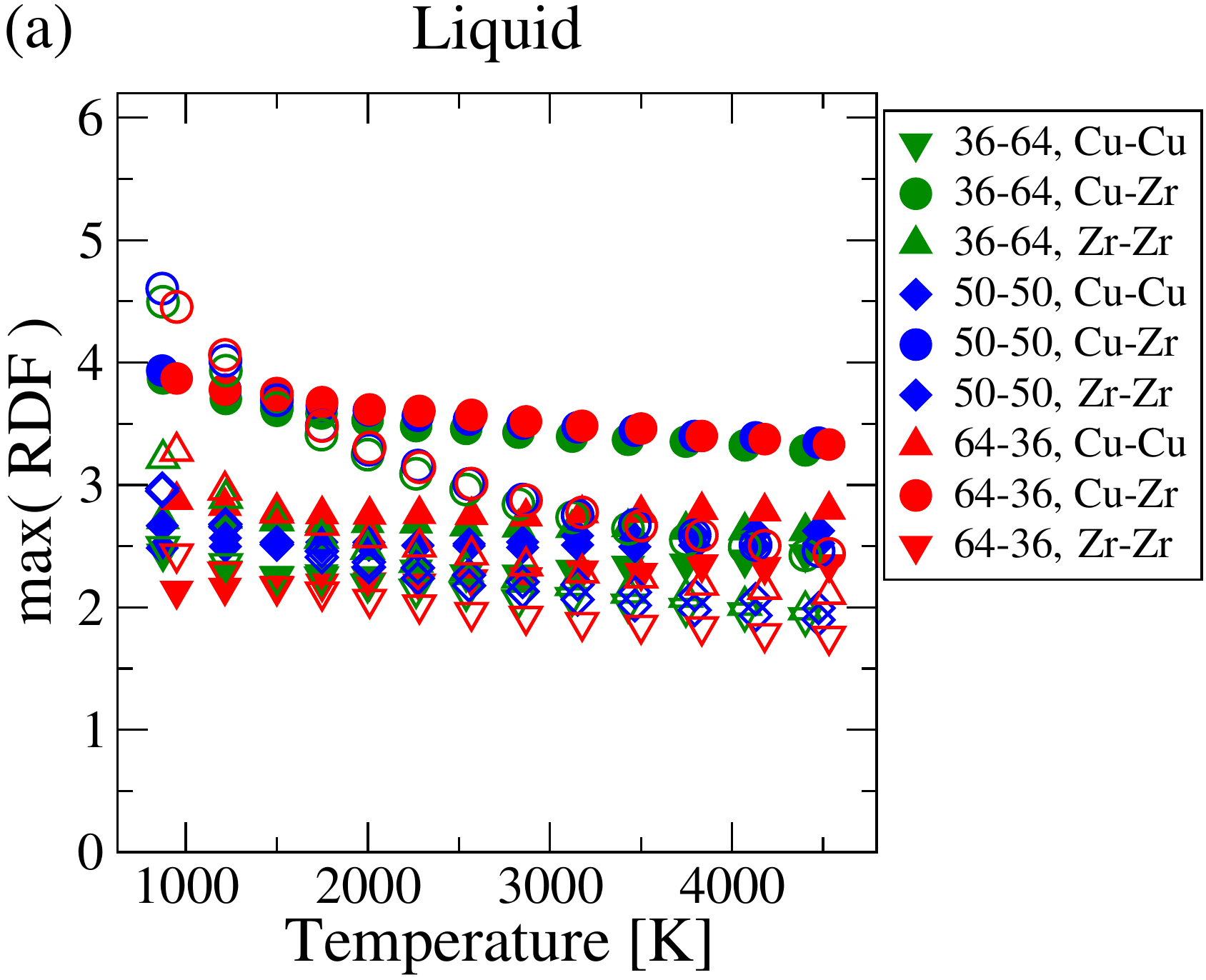}
	\includegraphics[width=5.8cm]{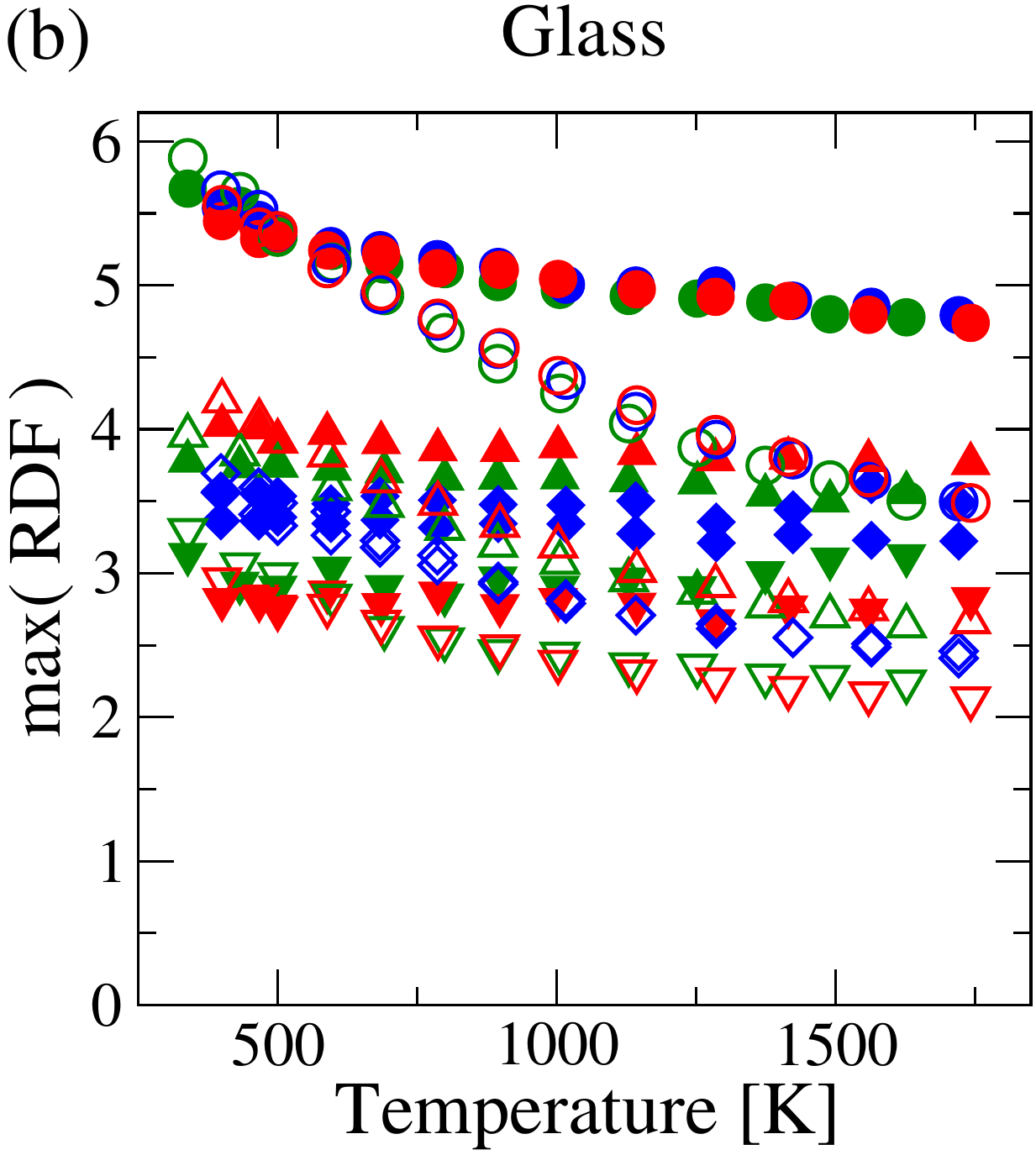}
	\caption{Maximum values of all RDFs of \fig{fig4} and \fig{fig5}, i.e., values of the RDFs at their first peak, with (a) showing results for the liquid and (b) for the glass. The full symbols represent isomorph data, the open symbols represent isochore data. While the isomorph values are not invariant, in particular for the Cu-Zr RDFs, for both the liquids and the glasses the general picture is that of a visibly better invariance along the isomorphs than along isochores of the same temperature variation.}
	\label{fig6}
\end{figure*}

Focusing on the height of the first RDF peak, \fig{fig6} shows these for all the data of \fig{fig4} and \fig{fig5}; for ease of comparison we included here also the analogous data for the glass-phase simulations (\Sec{glass}). The full symbols are the peak heights along the three isomorphs, whereas the open symbols are the peak heights along the corresponding isochores. Clearly, the variation is significantly larger along the isochores.

\begin{figure*}[t]
	\centering
	\includegraphics[width=0.24\linewidth]{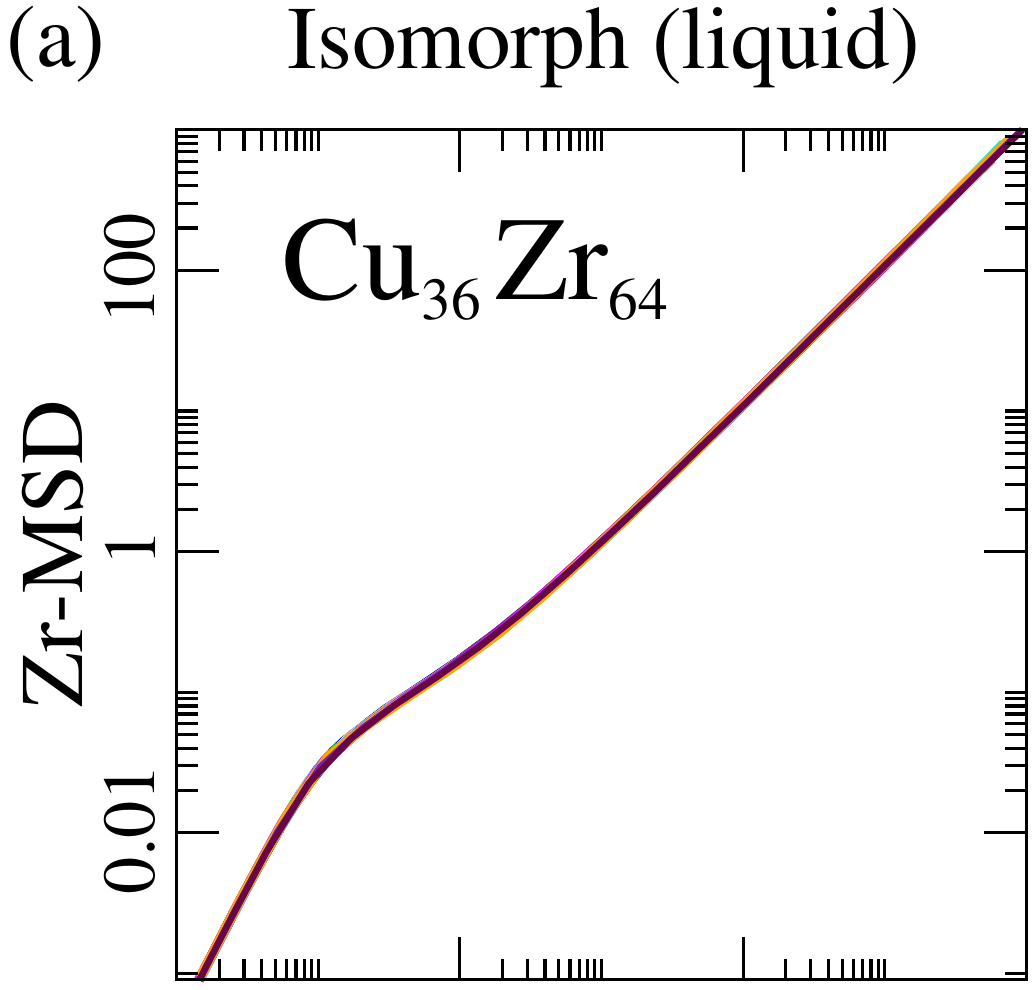}\hfill
	\includegraphics[width=0.24\linewidth]{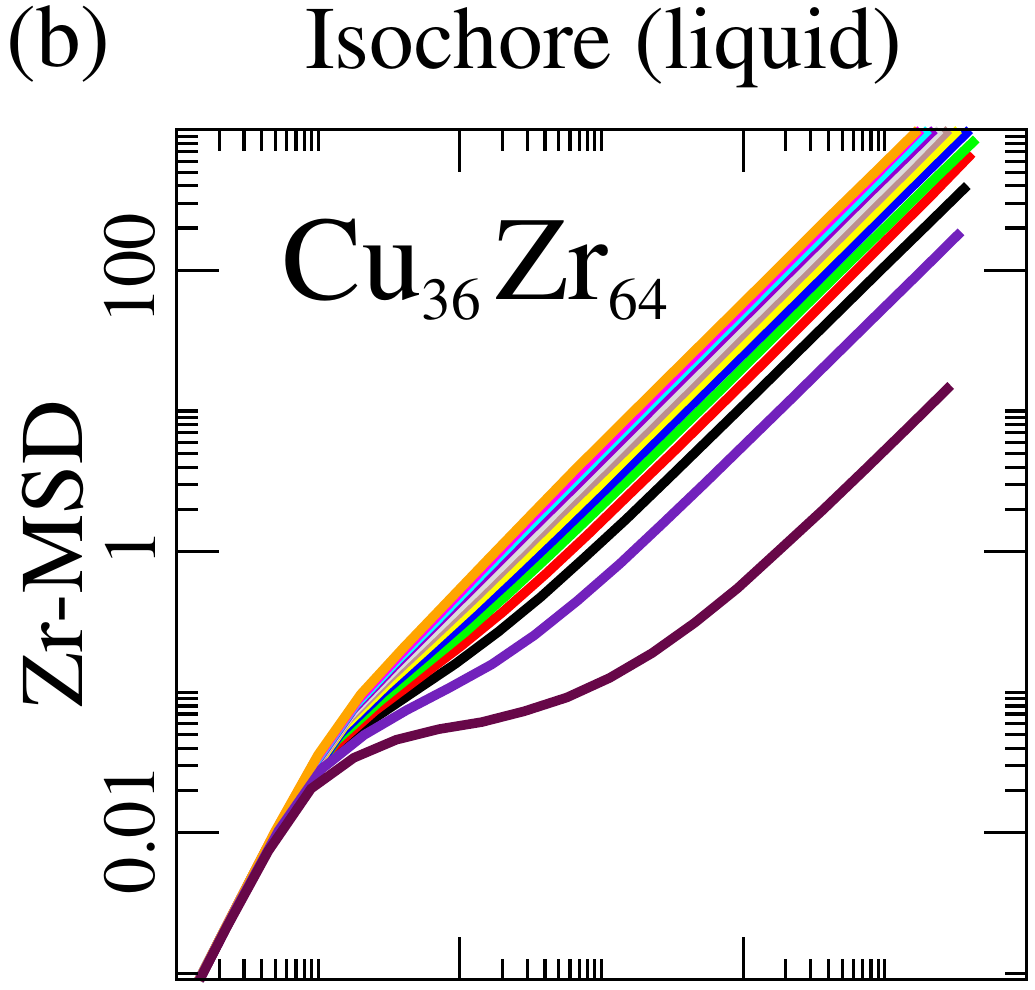}\hfill
	\includegraphics[width=0.24\linewidth]{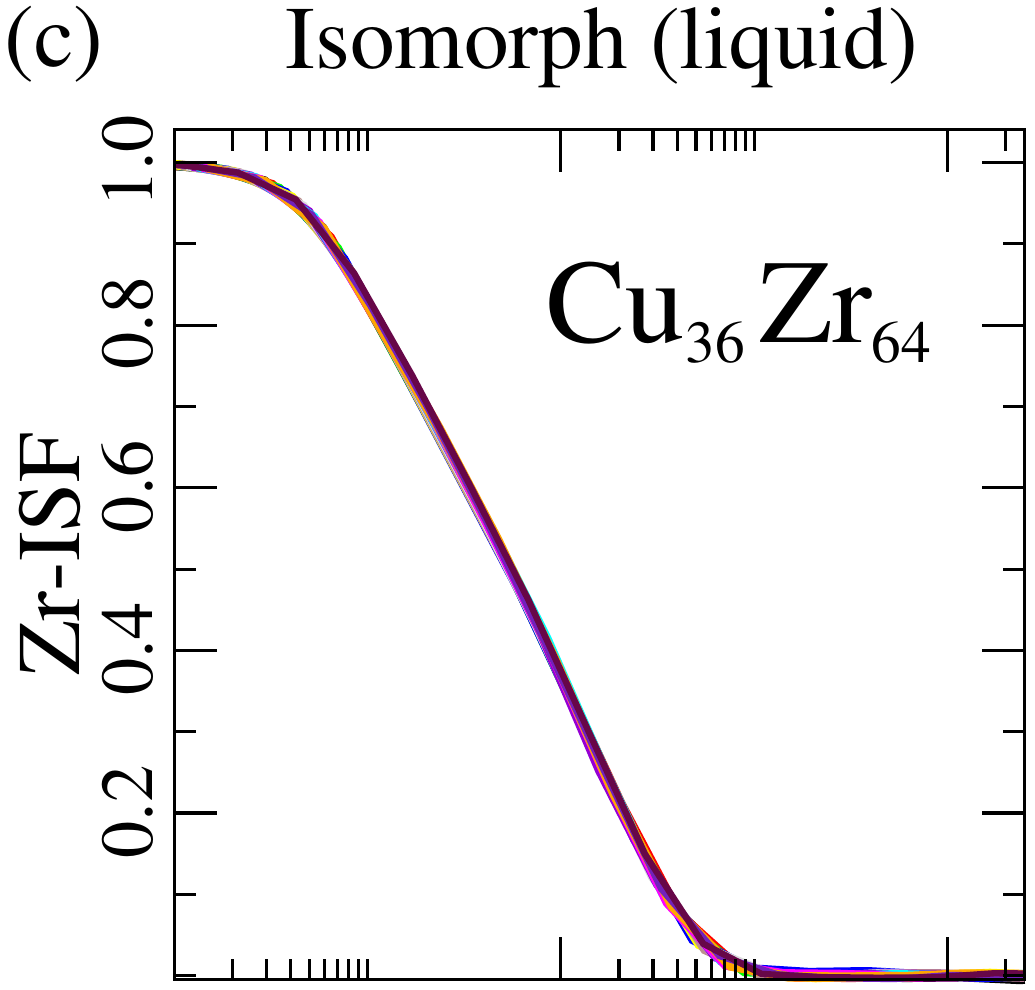}\hfill
	\includegraphics[width=0.24\linewidth]{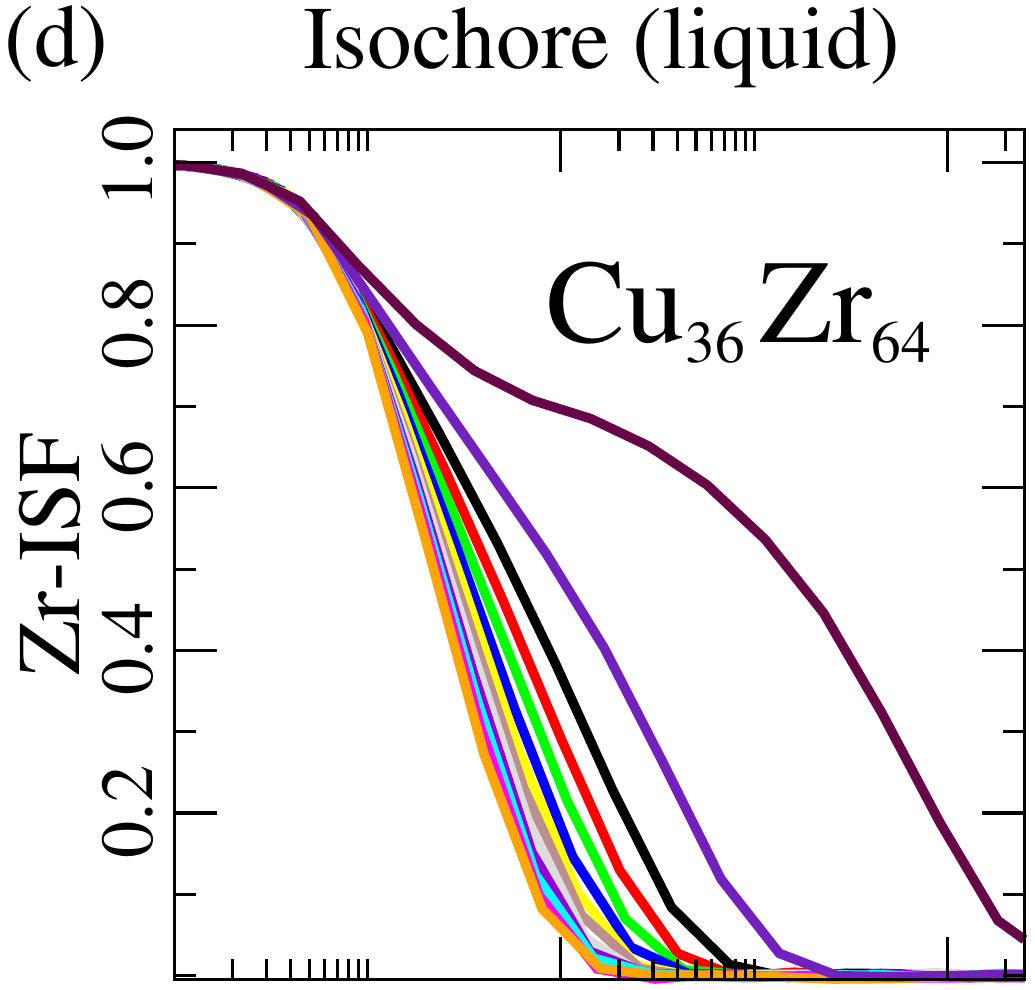}\hfill
	\\[1mm]
	\includegraphics[width=0.24\linewidth]{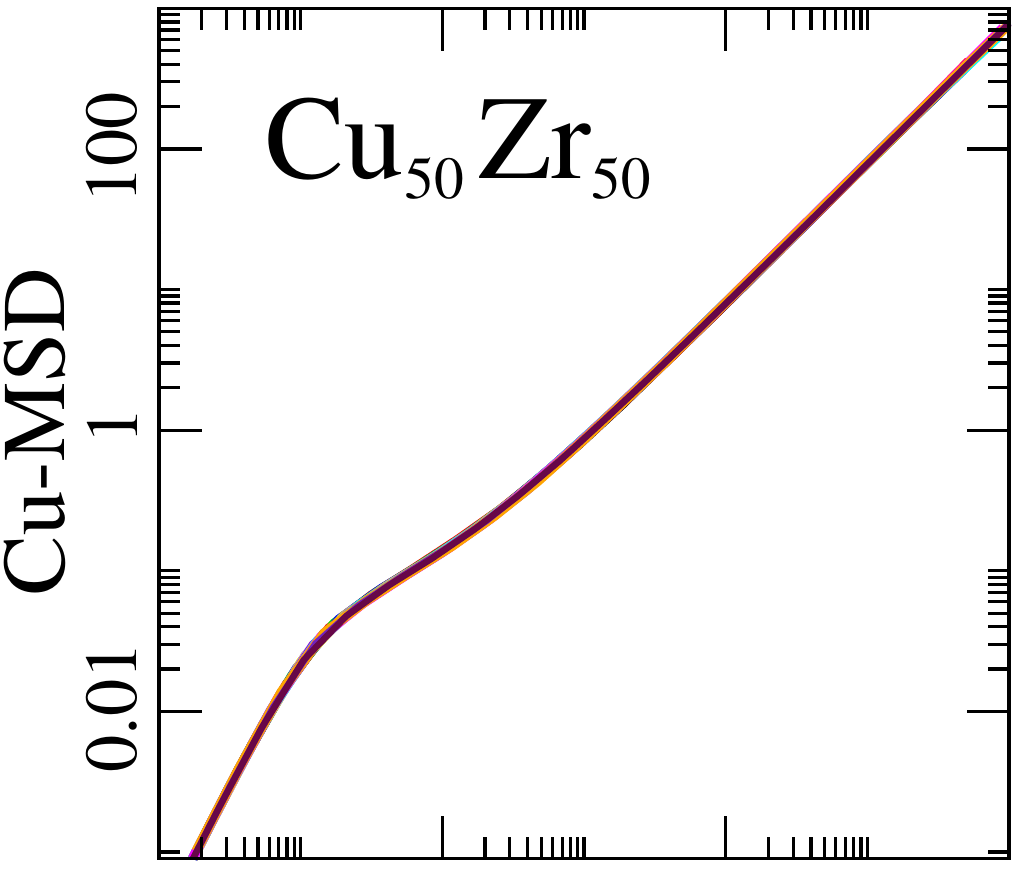}\hfill
	\includegraphics[width=0.24\linewidth]{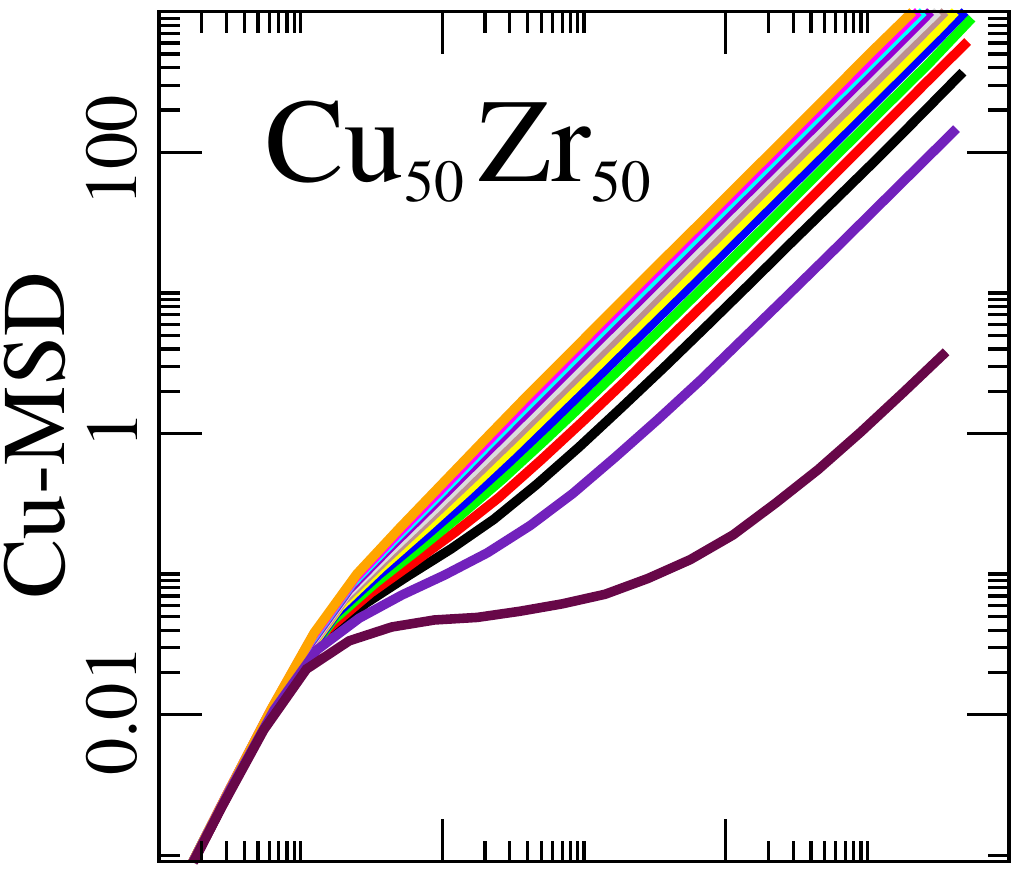}\hfill
	\includegraphics[width=0.24\linewidth]{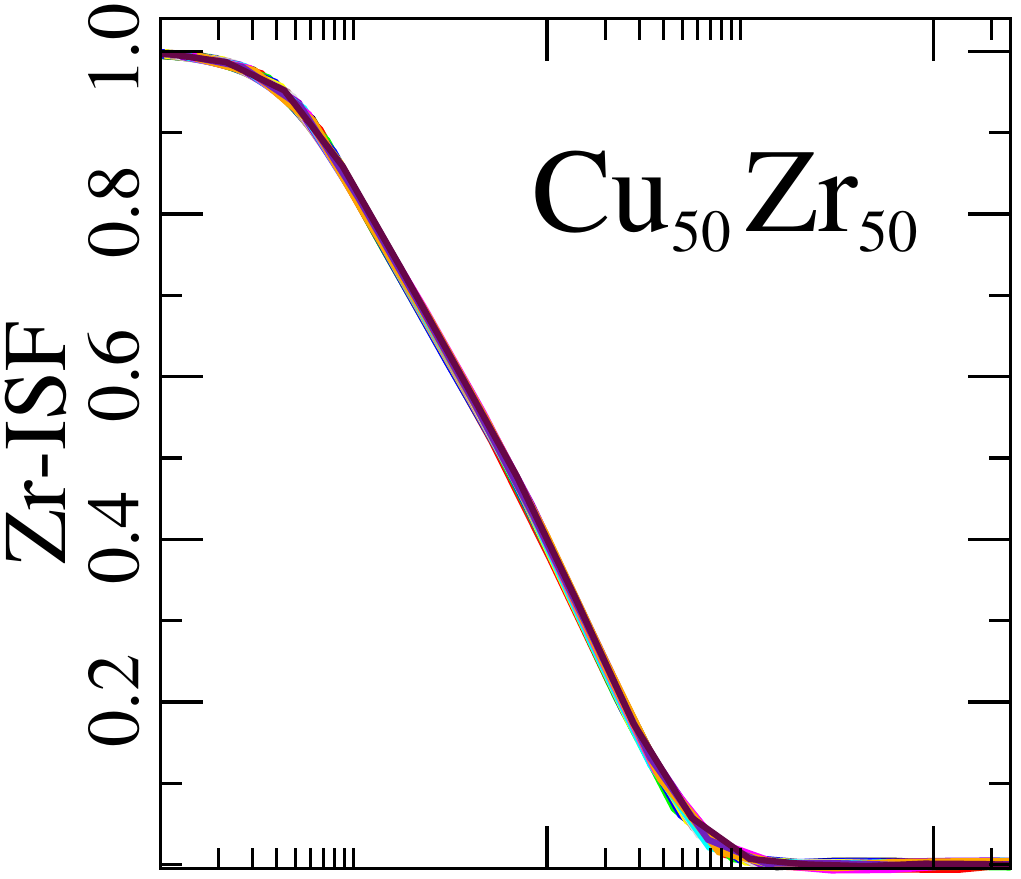}\hfill
	\includegraphics[width=0.24\linewidth]{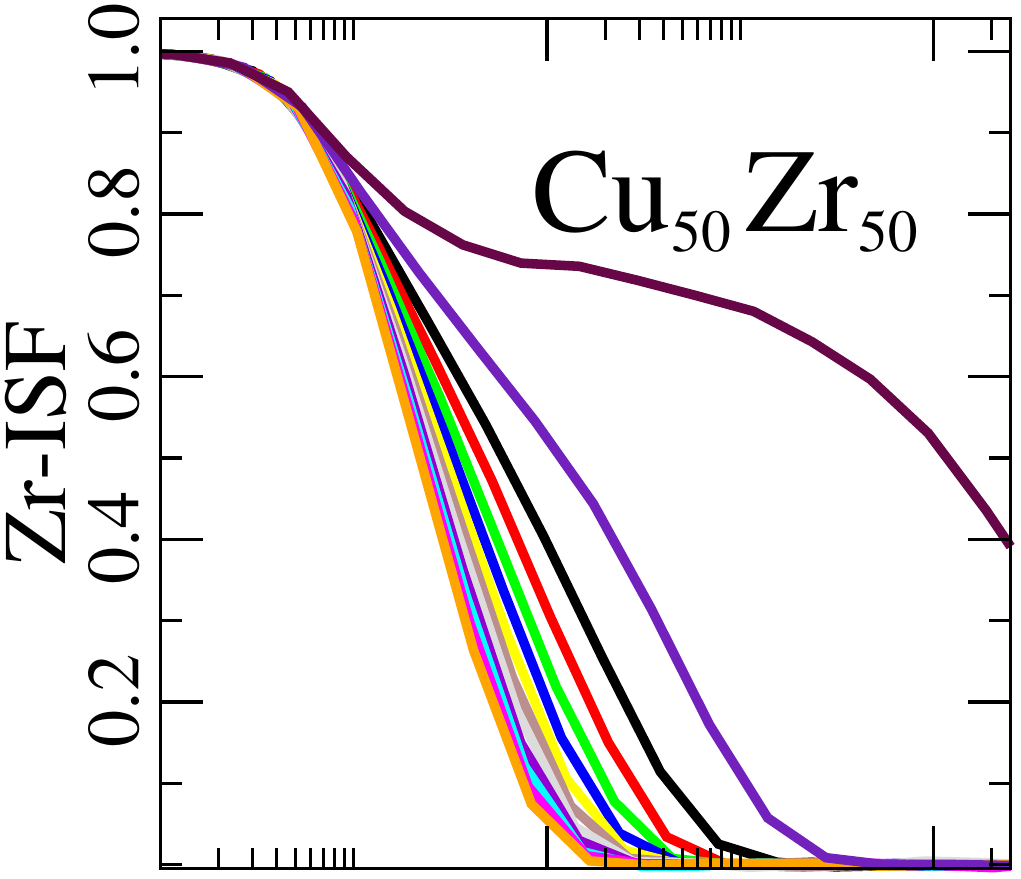}\hfill
	\\[1mm]
	\includegraphics[width=0.24\linewidth]{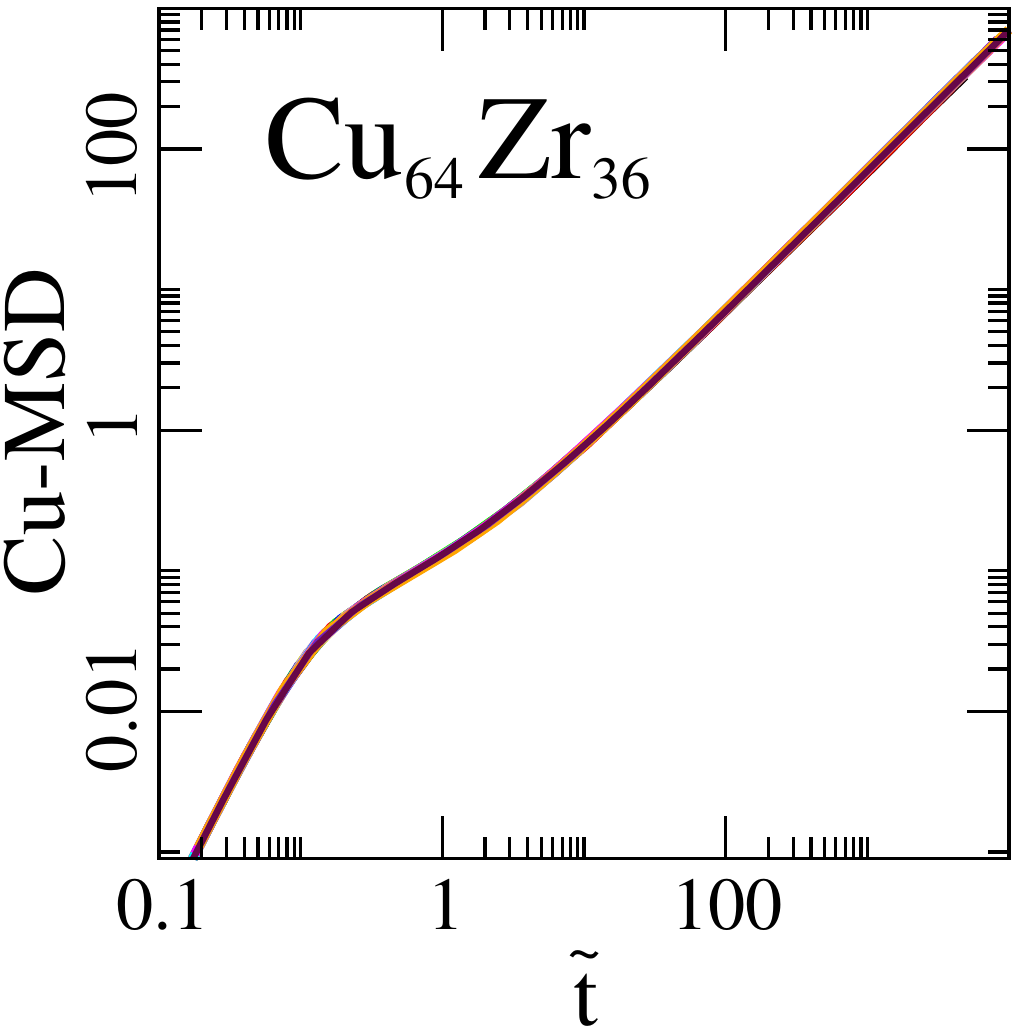}\hfill
	\includegraphics[width=0.24\linewidth]{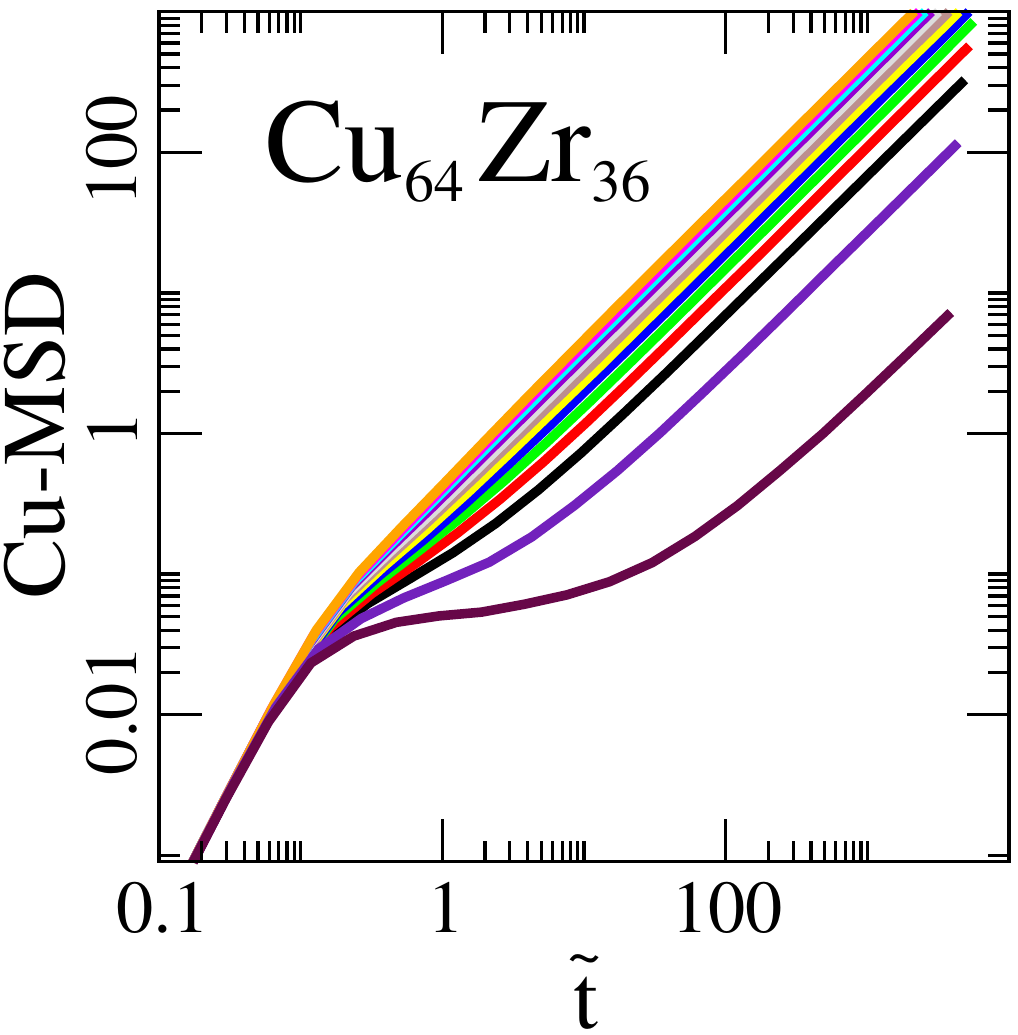}\hfill
	\includegraphics[width=0.24\linewidth]{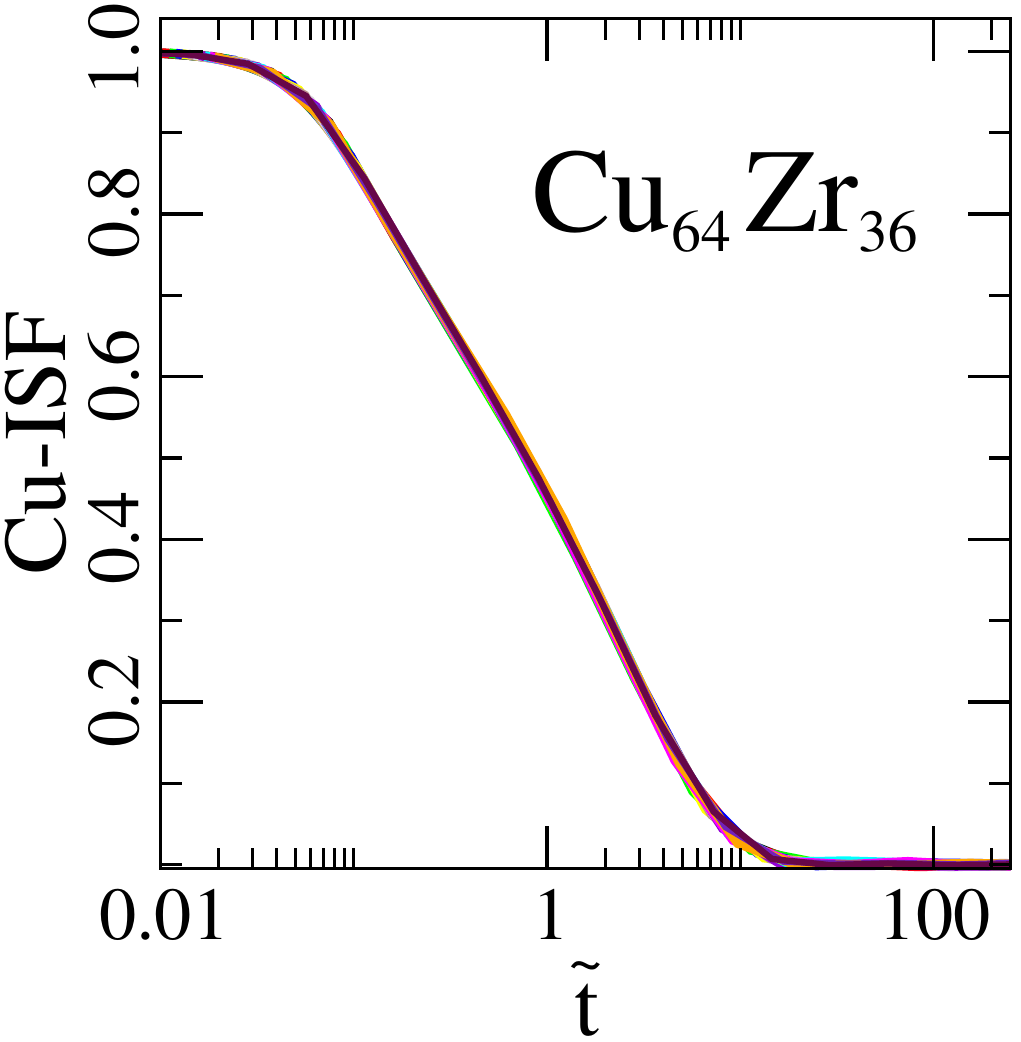}\hfill
	\includegraphics[width=0.24\linewidth]{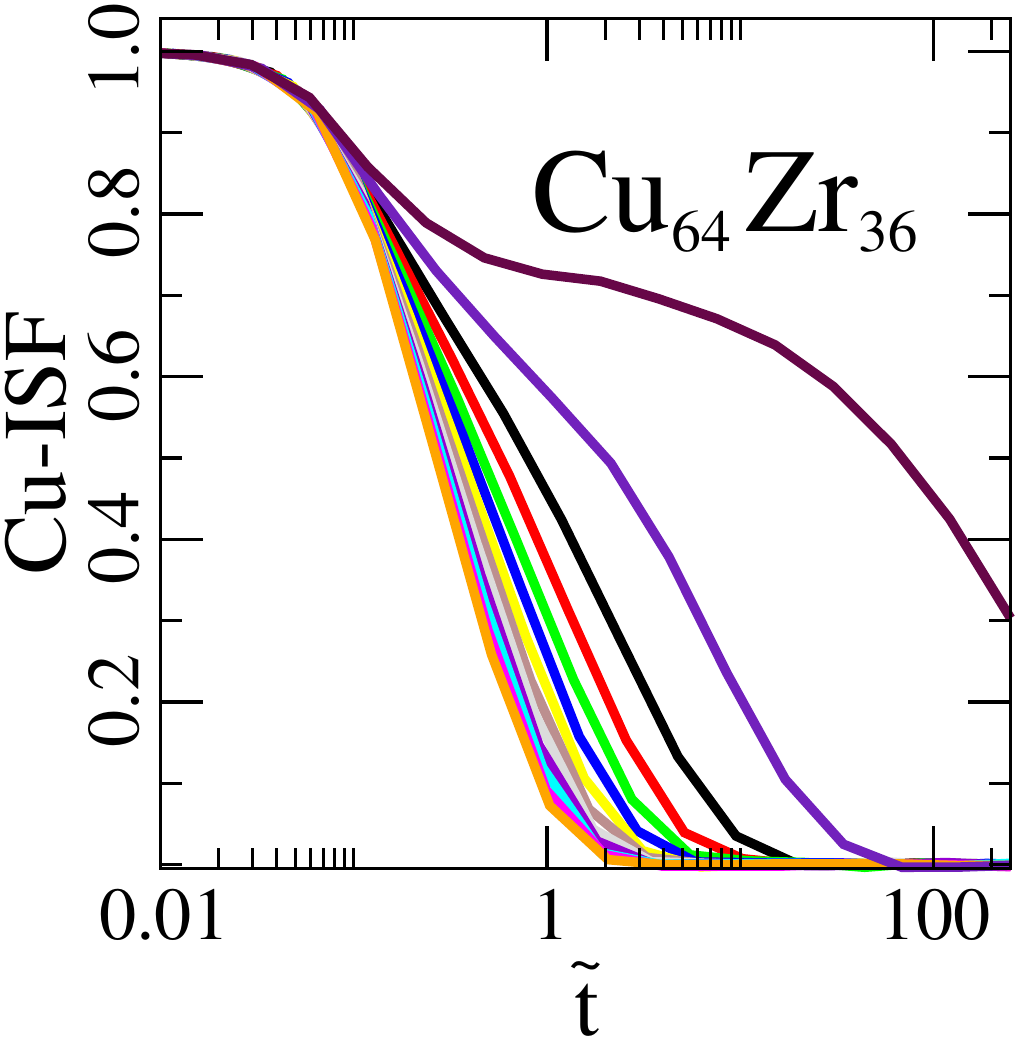}\hfill
	\\[1mm]
	\caption{Dynamics along the isomorphs and isochores of the liquids. (a) and (b) give data for the reduced MSD as a function of the reduced time, (c) and (d) give the reduced-unit intermediate incoherent scattering function at the wave vector  $2\pi\rho^{1/3}$ (which is constant in reduced units). There is good isomorph invariance of the dynamics, but a significant variation along the isochores (except in the short-time ballistic region where the reduced MSD by definition is $3\tilde{t}^2$ at all state points).}
	\label{fig7}
\end{figure*}

Next we investigated the liquid-phase dynamics. \Fig{fig7} shows results for the reduced-unit mean-square displacement (MSD) along isomorphs and isochores, respectively (two left columns). We focus on the majority atom MSD, but found that data for the minority atom are entirely similar (not shown). Clearly, the MSD is isomorph invariant and varies significantly along the isochores. Note that the short-time ballistic-region collapse seen in all cases follows from the definition of reduced units, i.e., this collapse applies throughout the phase diagram of any system. The two right-hand columns of \fig{fig7} show similar data for the incoherent intermediate scattering function (evaluated at the wave vector $2\pi\rho^{1/3}$ that is constant in reduced units). Again, isomorph invariance is clearly demonstrated.

Returning to \fig{fig6}, in view of \fig{fig7} one may ask: which structural features are most important for the dynamics? \Fig{fig6} shows that the majority component self-RDF shows the best isomorphic scaling. This suggests that the dynamics of all atoms are largely determined by the majority species, i.e., that the two non-equimolar mixtures act as effective one-component systems.

\section{Structure and dynamics in the glass phase}\label{glass}

The above investigation was repeated in the glass phase of the three mixtures. Isomorph theory is traditionally formulated by reference to thermal equilibrium \cite{IV,sch14,dyr14}, but we ignored this and proceeded pragmatically as if a glass were an equilibrium system. Each of the three glasses were prepared by cooling with a constant rate from a configuration at the reference state point ($T=1500$K) to temperature 500 K. For each composition, once a glass configuration was obtained at the reference state point, we generated an isomorph in the same way as the liquid isomorphs by repeated DICs involving 5\% density changes. Again, for comparison we probed the RDF and the dynamics also at isochoric state points with the same temperature variation as that of the isomorphs (compare \fig{fig1}). 

\begin{figure*}[t]
	\centering
	\includegraphics[width=0.25\linewidth]{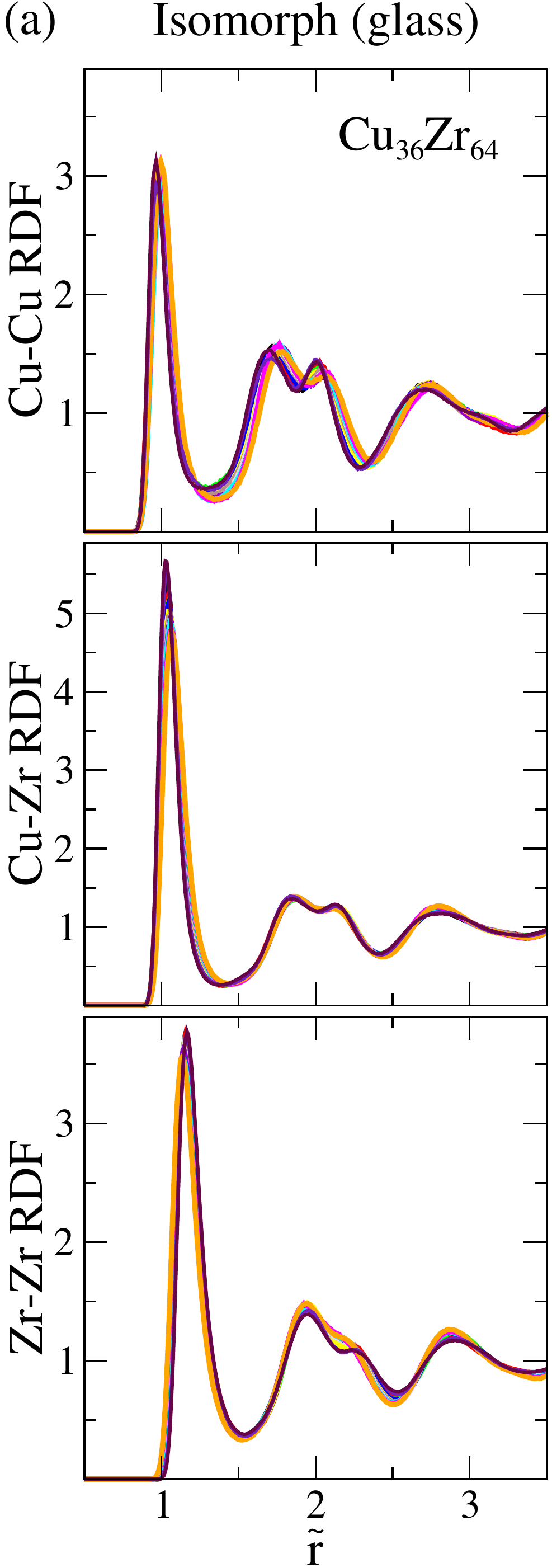}\hfill
	\includegraphics[width=0.25\linewidth]{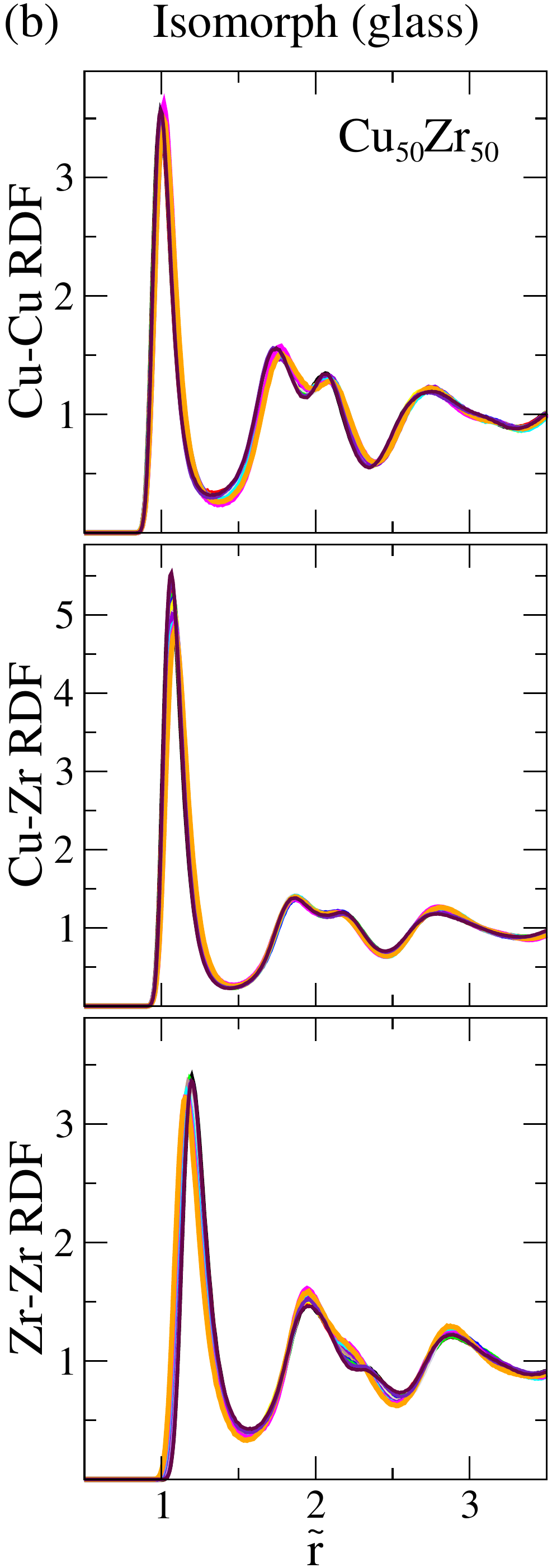}\hfill
	\includegraphics[width=0.25\linewidth]{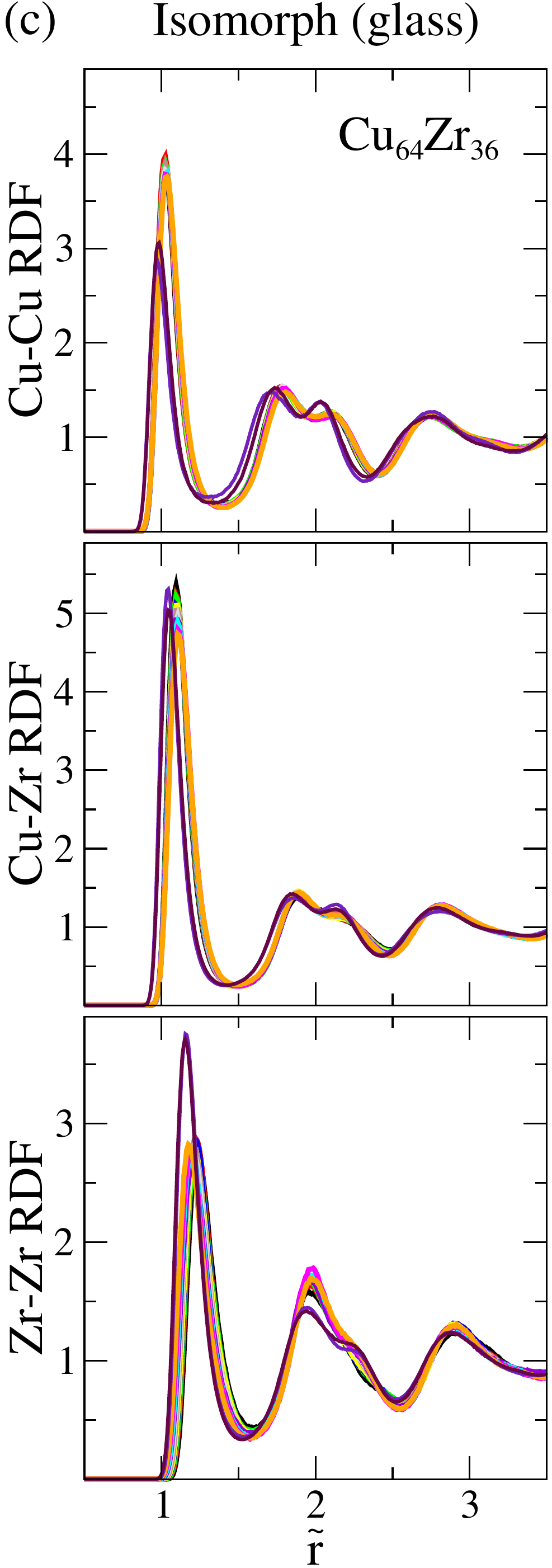}
	\caption{Isomorph glass RDFs plotted as a function of the reduced pair distance $\tilde{r}$. (a), (b), and (c) give reduced-unit RDFs along an isomorph for each of the compositions studied. The picture is pretty much the same as in the liquid phase, with good isomorph invariance except for the somewhat more noisy data that are  presumably due to the fact that, basically, only a single configuration and its vibrations is probed. Significant deviations from isomorph invariance are observed, though, for the 64\% Cu mixture, both at the highest densities for the Cu-Cu RDF and at the lowest densities for the Zr-Zr RDF.}
	\label{fig8}
\end{figure*}

\begin{figure*}[t]
	\centering
	\includegraphics[width=0.25\linewidth]{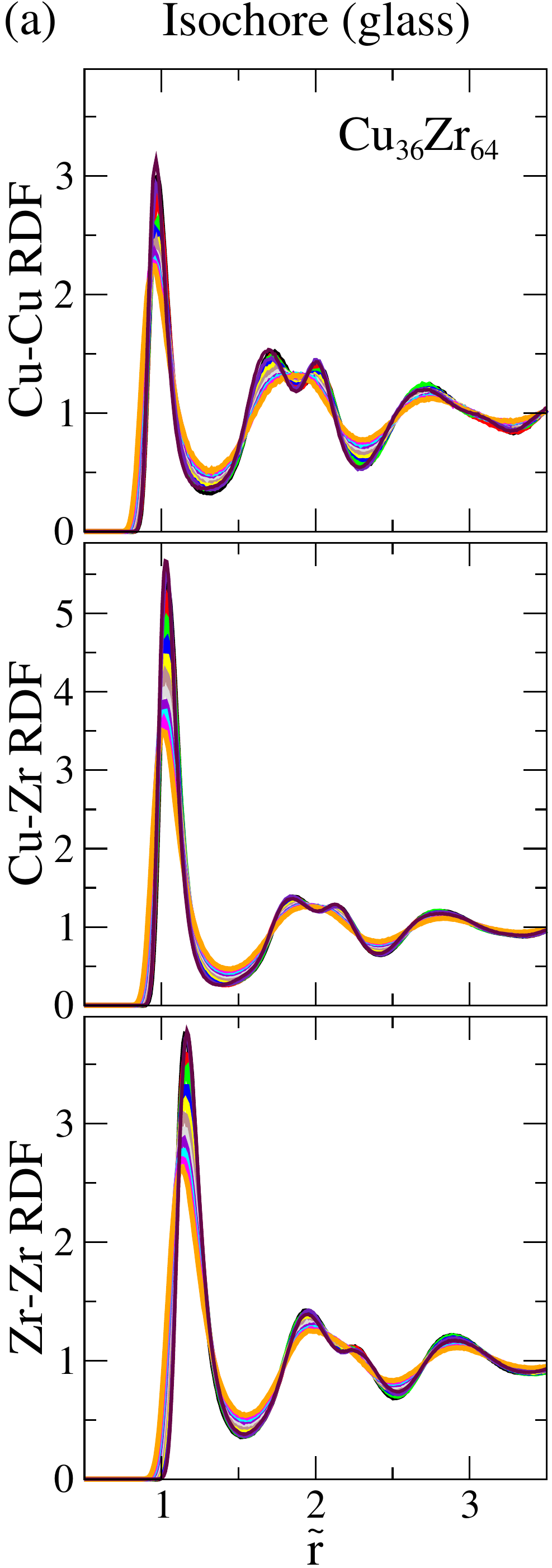}\hfill
	\includegraphics[width=0.25\linewidth]{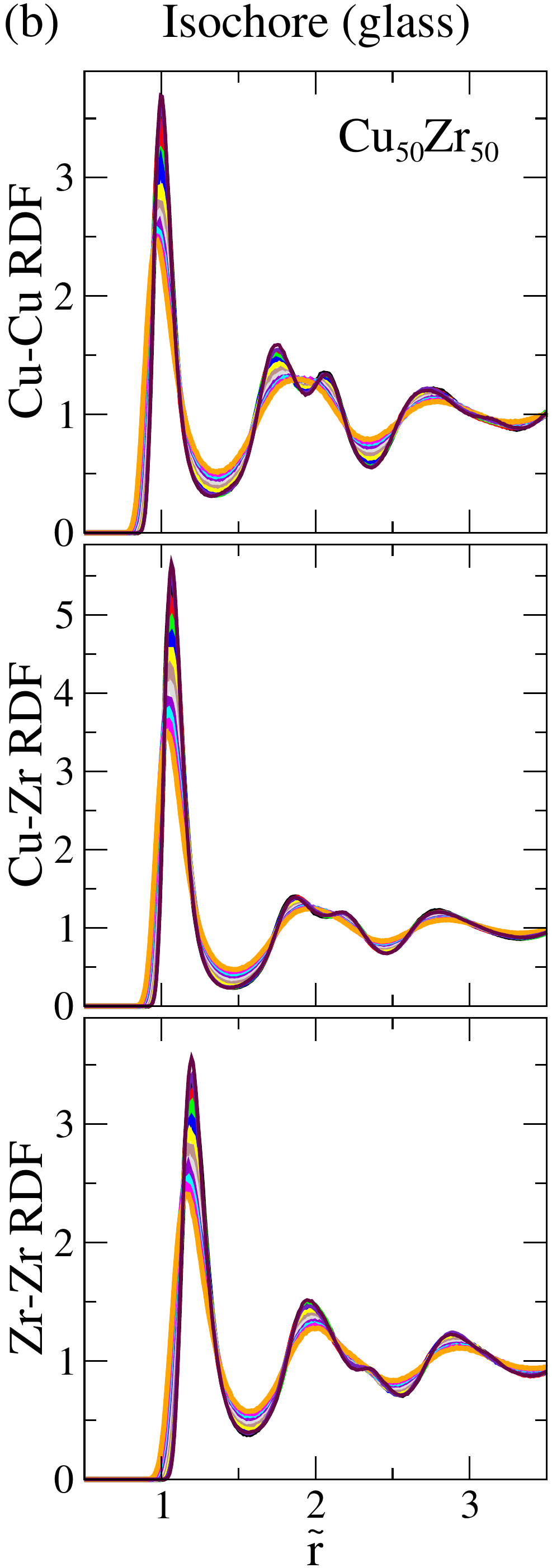}\hfill
	\includegraphics[width=0.25\linewidth]{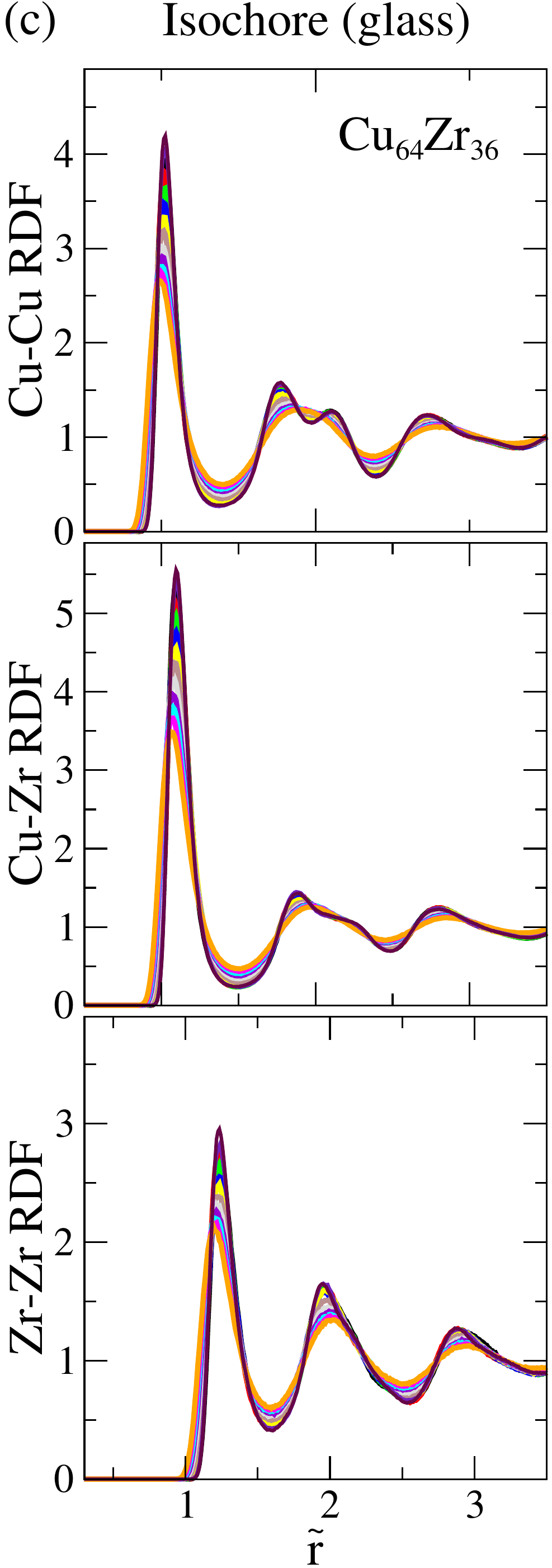}
	\caption{Isochore glass-state RDFs, plotted as in \fig{fig8}, showing a considerably larger variation of the structure.}
	\label{fig9}
\end{figure*}

The RDFs are shown in \fig{fig8} (isomorphs) and in \fig{fig9} (isochores). The picture is similar to that of the liquid phase: overall, good invariance along the isomorphs is seen in contrast to a substantially larger variation along the isochores. This is also the conclusion from \fig{fig6} showing all the first-peak heights as a function of the temperature.

\begin{figure*}[t]
	\centering
	\includegraphics[width=0.24\linewidth]{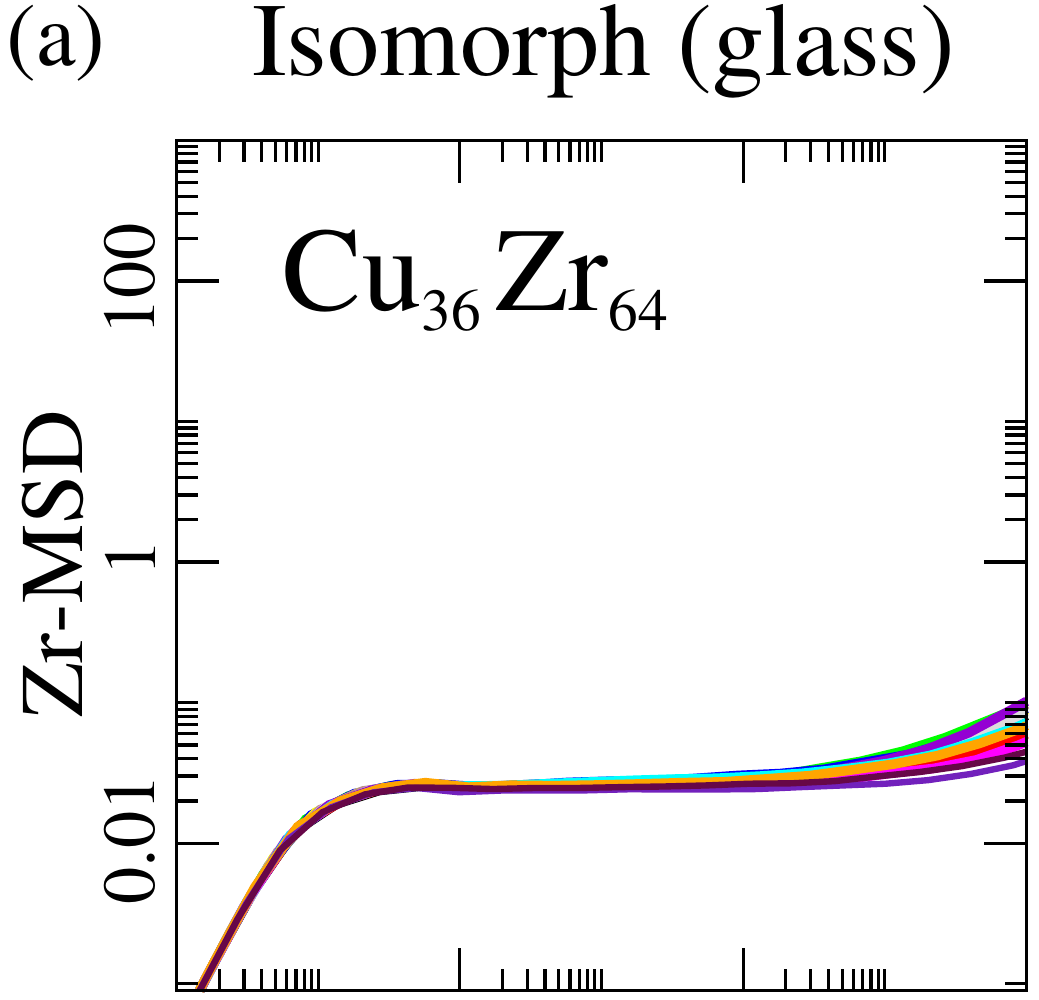}\hfill
	\includegraphics[width=0.24\linewidth]{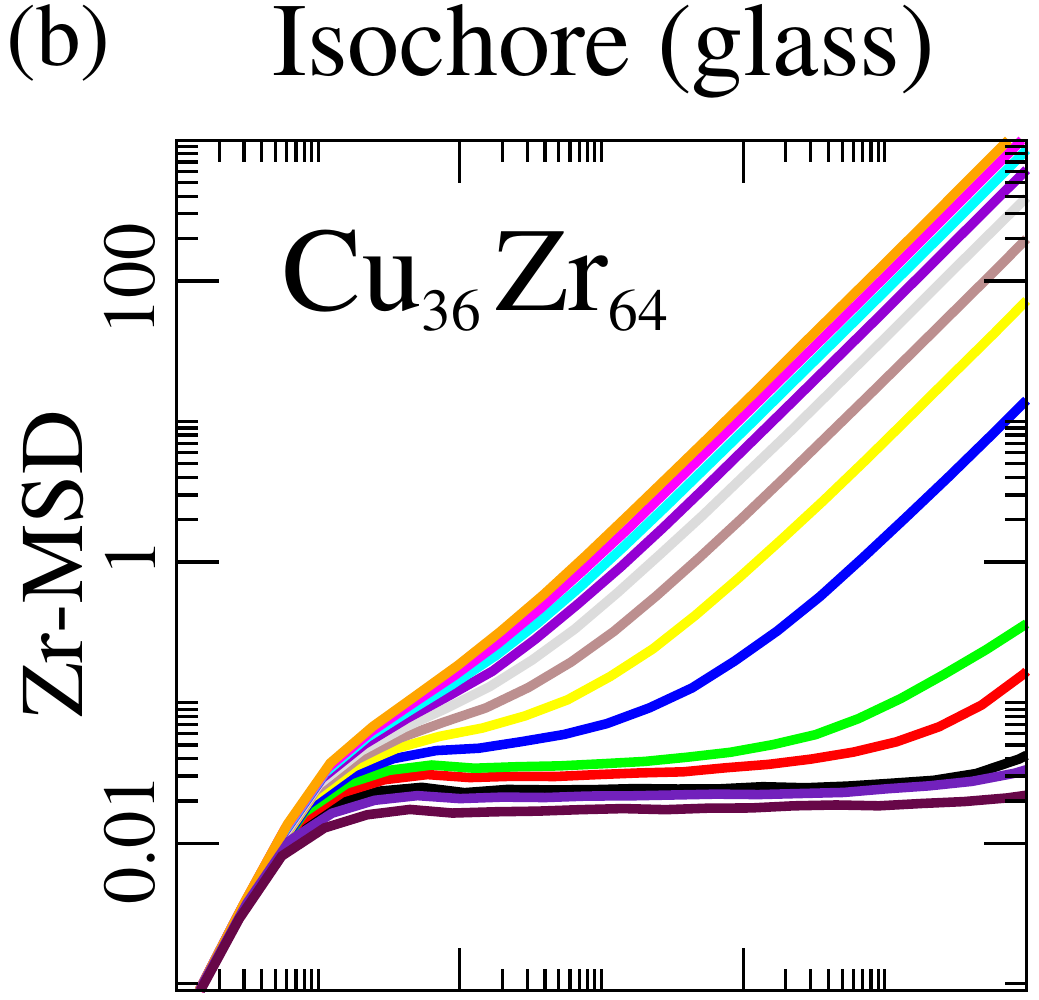}\hfill
	\includegraphics[width=0.24\linewidth]{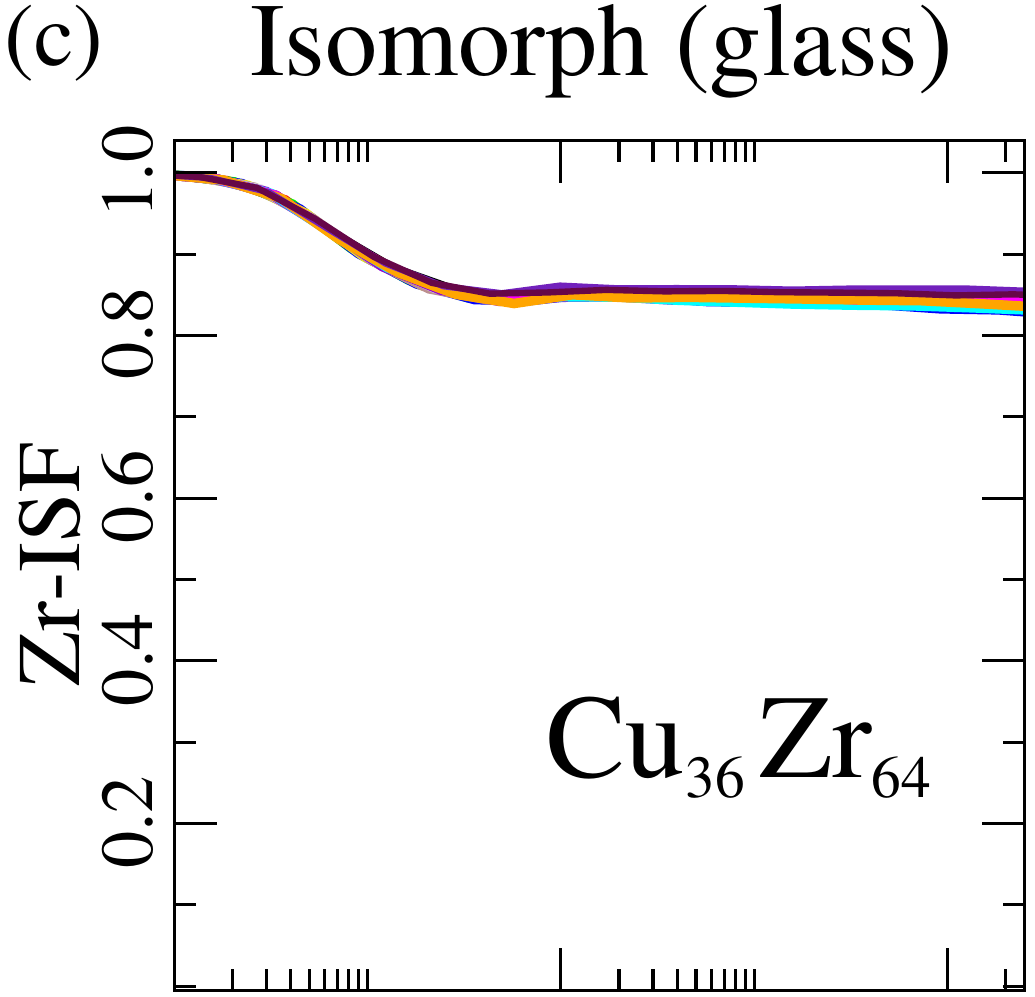}\hfill
	\includegraphics[width=0.24\linewidth]{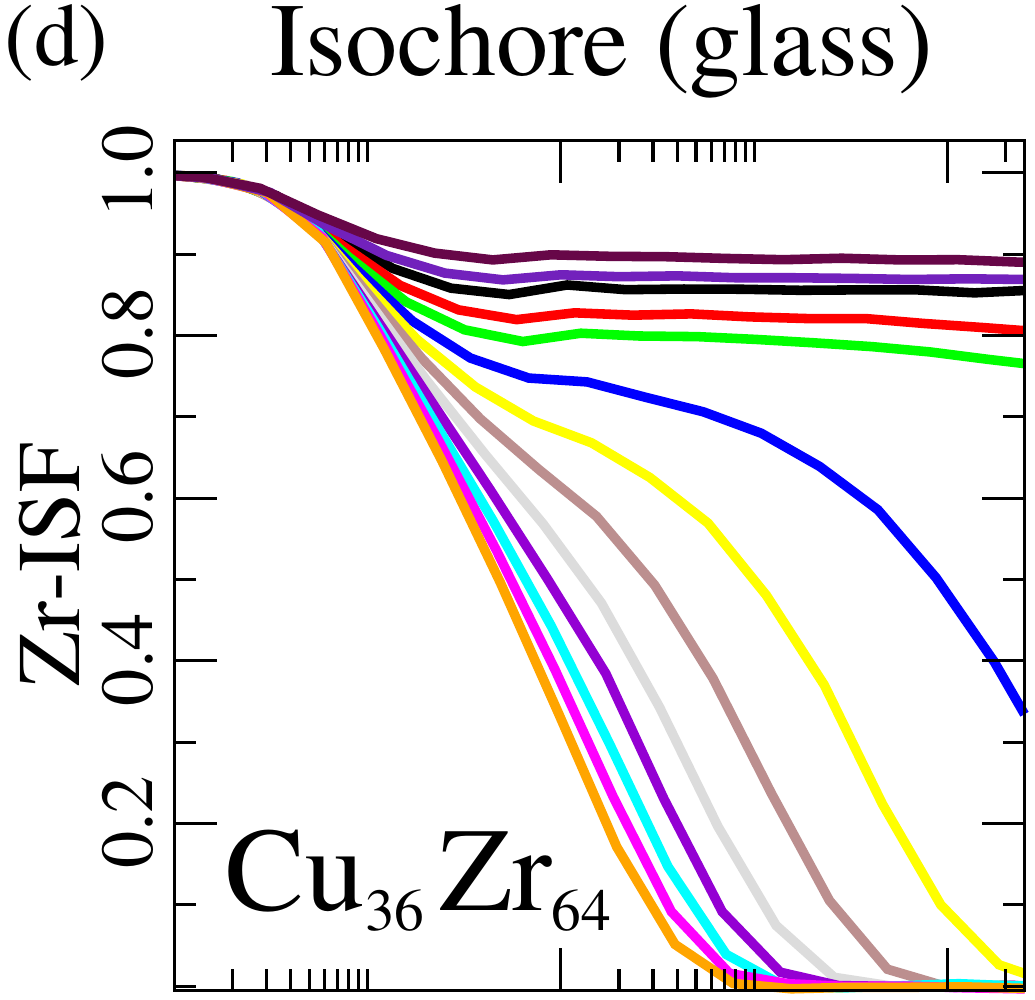}\hfill
	\\[1mm]
	\includegraphics[width=0.24\linewidth]{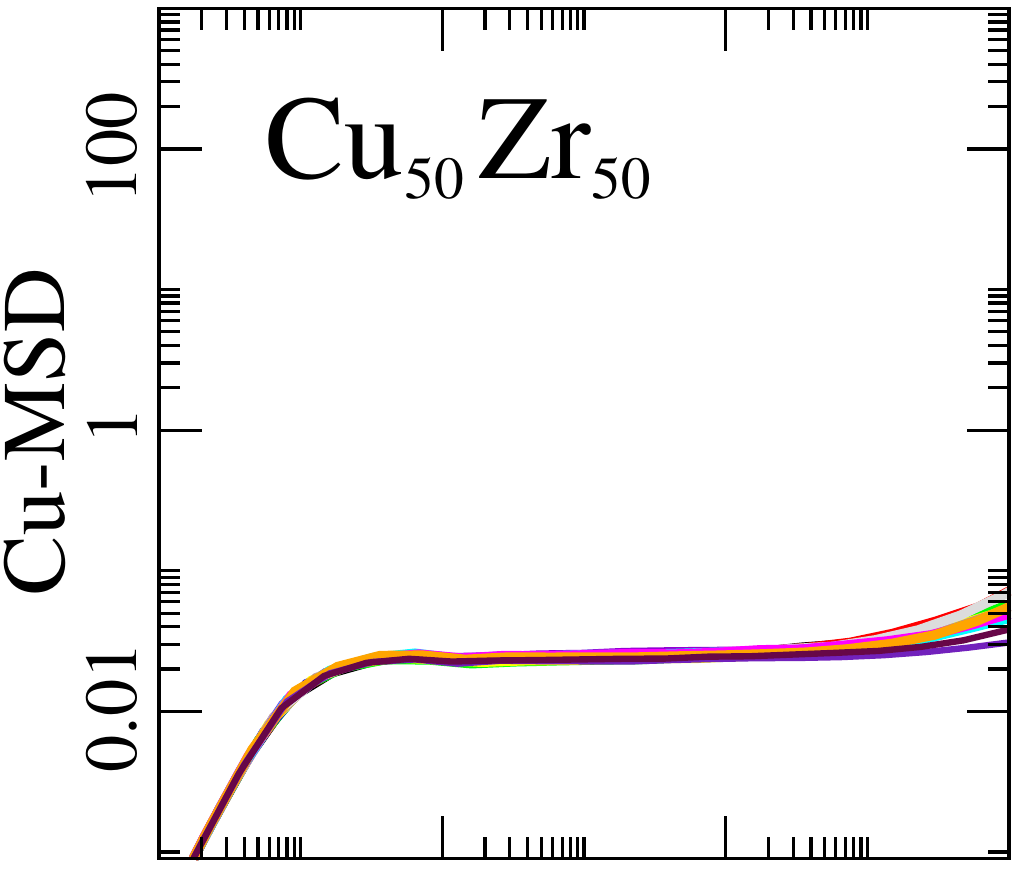}\hfill
	\includegraphics[width=0.24\linewidth]{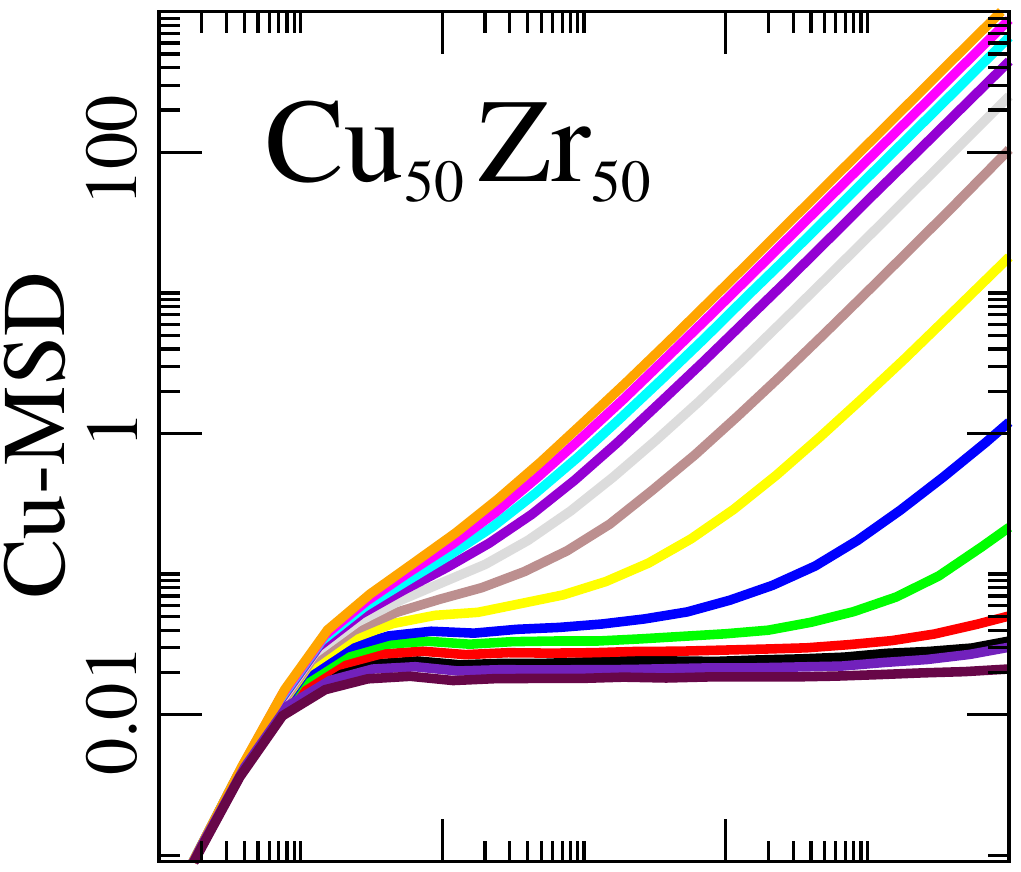}\hfill
	\includegraphics[width=0.24\linewidth]{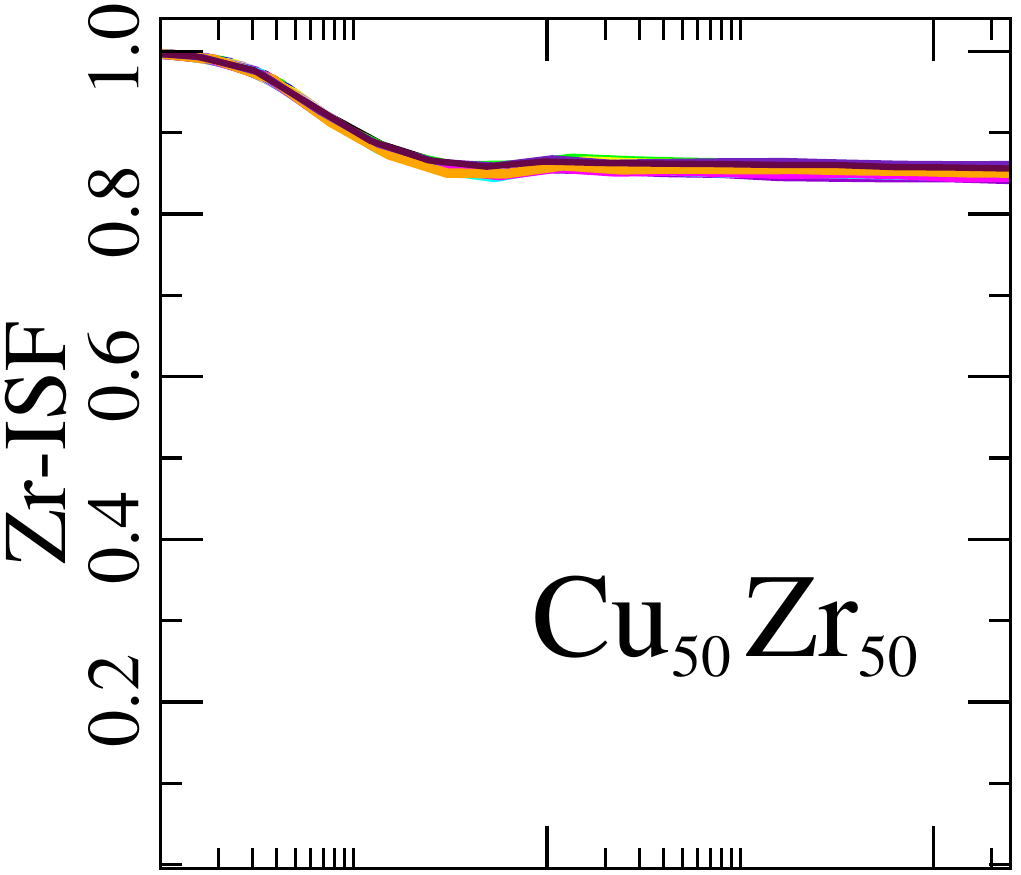}\hfill
	\includegraphics[width=0.24\linewidth]{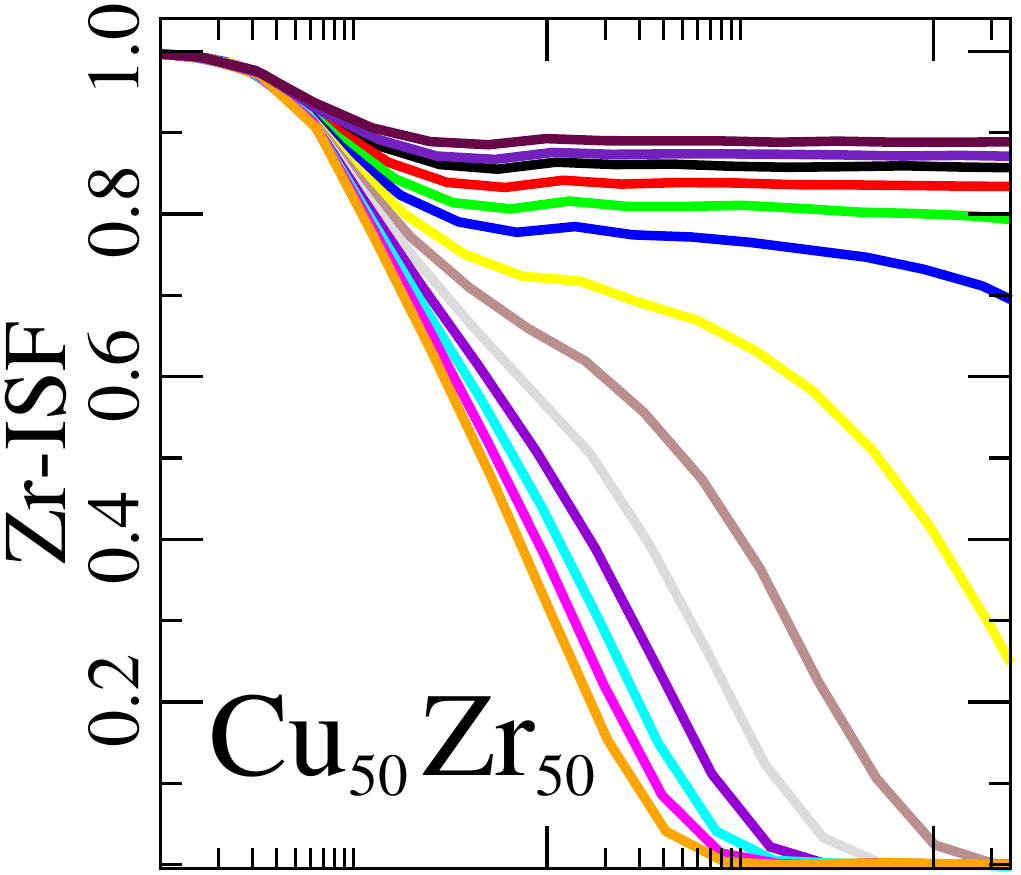}\hfill
	\\[1mm]
	\includegraphics[width=0.24\linewidth]{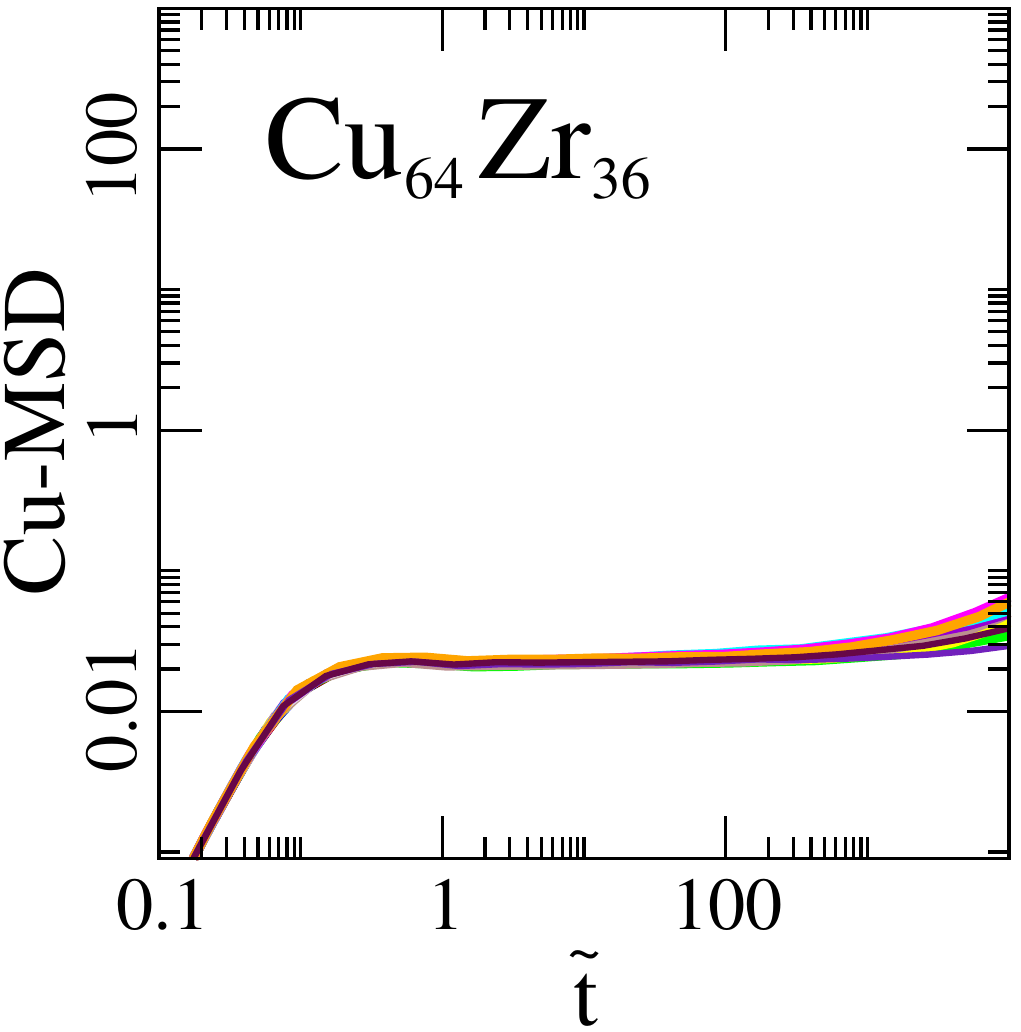}\hfill
	\includegraphics[width=0.24\linewidth]{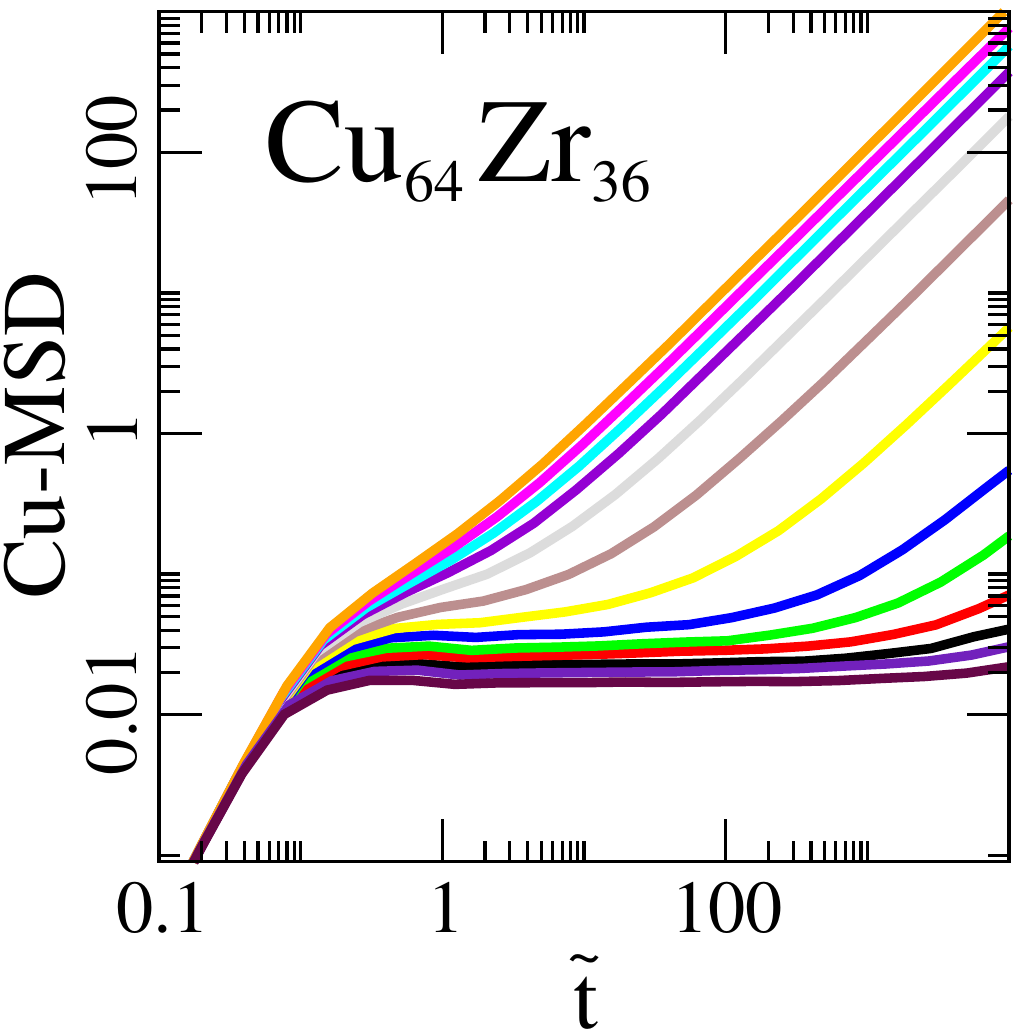}\hfill
	\includegraphics[width=0.24\linewidth]{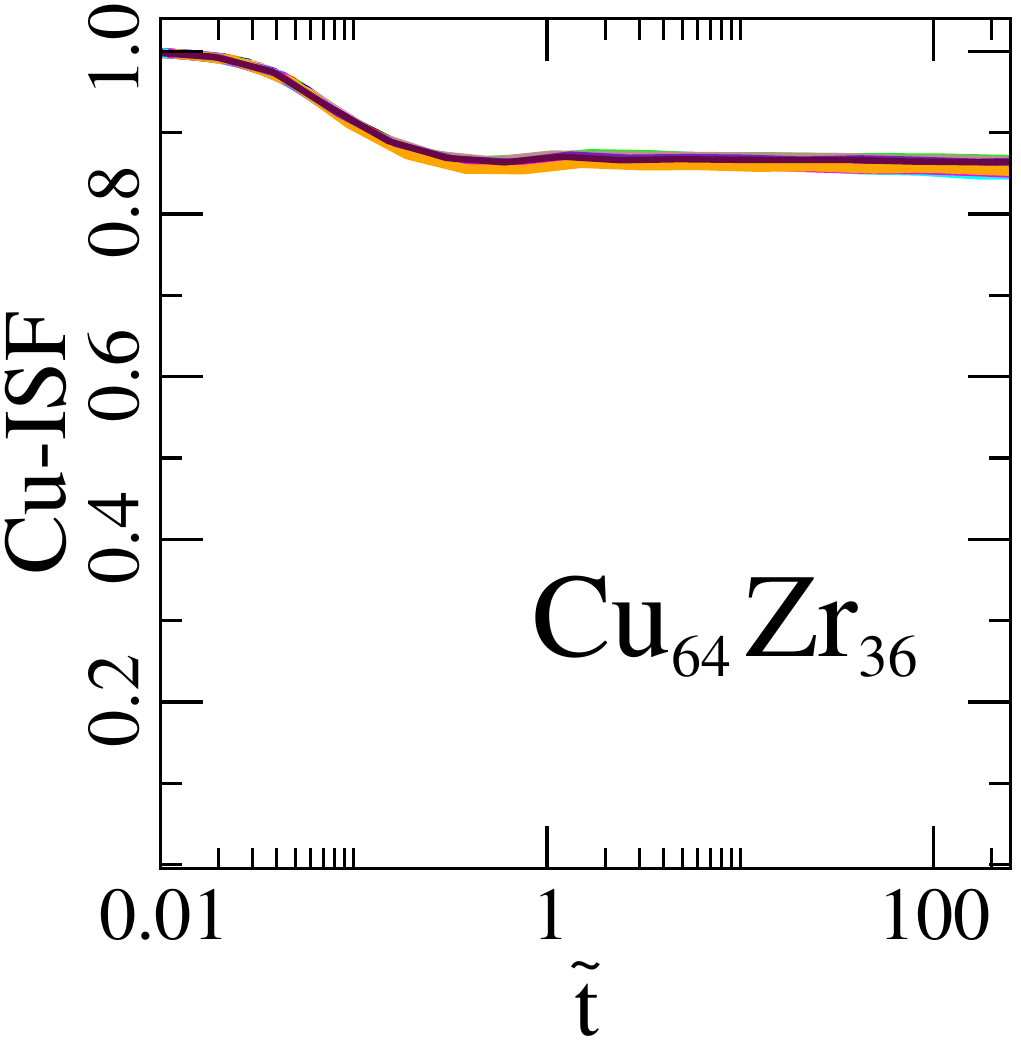}\hfill
	\includegraphics[width=0.24\linewidth]{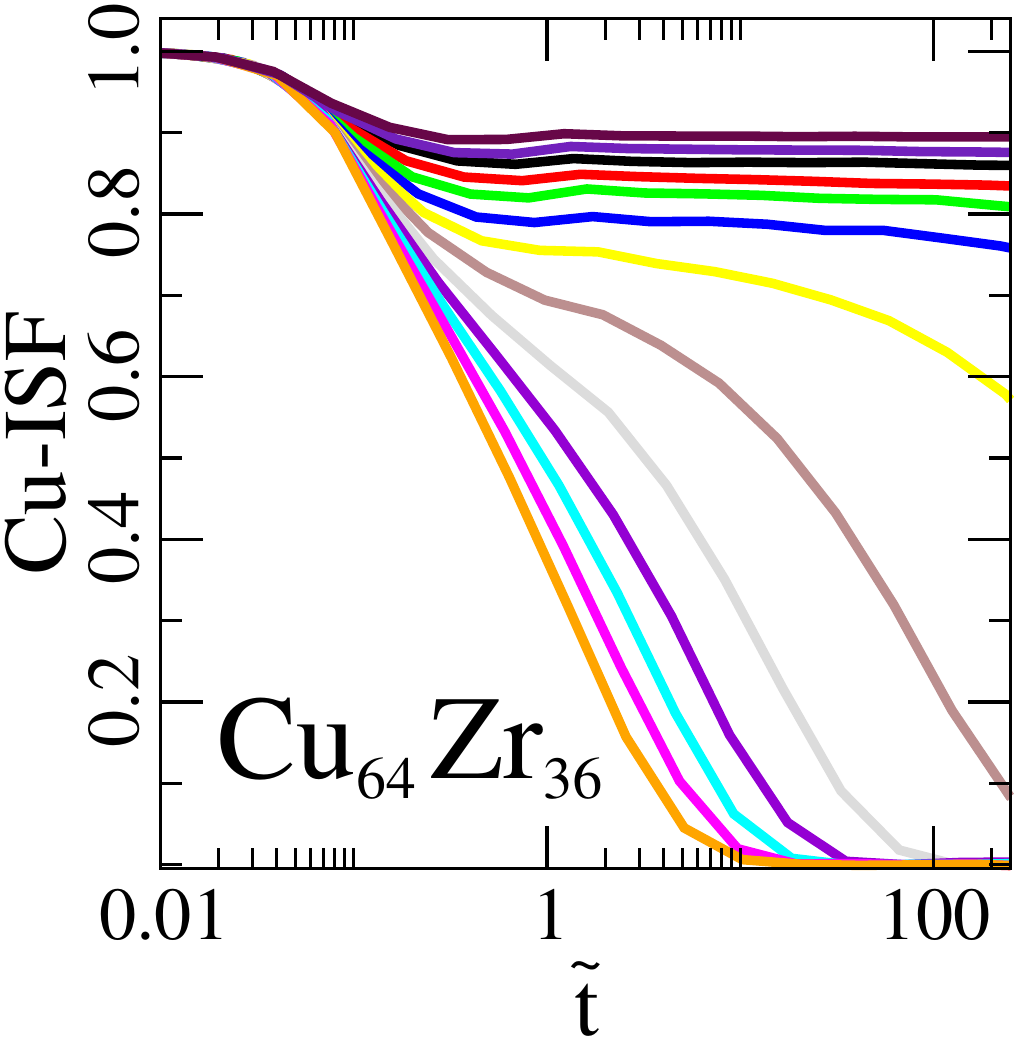}\hfill
	\caption{Dynamics along the glass isomorphs and isochores of the glasses. (a) and (b) give data for the reduced MSD as a function of the reduced time, (c) and (d) give the reduced-unit intermediate incoherent scattering function data at the wave vector  $2\pi\rho^{1/3}$. Because the system is a glass, along the isomorphs the MSD is constant over a very long time and, correspondingly, the incoherent intermediate scattering function does not decay to zero. For the isochore state points, raising the temperature takes the system closer to a liquid with a MSD that at long times is proportional to time and an intermediate incoherent scattering function that decays to zero at long times. The overall picture is that, as for the liquid (\fig{fig7}), there is good isomorph invariance of the dynamics, but a significant variation along the isochores.}
	\label{fig11}
\end{figure*}

The dynamics of the glasses is investigated via the MSD and the incoherent intermediate structure factor in \fig{fig11}. A glass consists mostly of atoms frozen at fixed positions, merely vibrating there. Thus the MSD is virtually constant, though some particle motion is discernible at the longest times. This ``glass flow'' motion does not appear to be isomorph invariant, but we found no systematic variation with the density. This indicates that the non-invariance reflects statistical uncertainty. Along the isochores (second column), it is clear that the glass gradually melts as temperature is increased when moving from the black curve representing the 500 K reference state point. The fact that the particles in the glass virtually do not move except for vibrations is also visible in the incoherent intermediate scattering function (right two columns of \fig{fig11}), which along the isomorphs stabilize at a constant level at long times. In contrast, many of the isochore curves go to zero at long times, reflecting an ease of motion with increasing temperature that is not achieved along the isomorphs. 

As mentioned, a glass is an out-of-equilibrium system, i.e., not a typical member of a canonical-ensemble distribution. It may be surprising that one can ignore this and go ahead by constructing isomorphs by the direct isomorph check, isomorphs that turn out to work basically just as well as the equilibrium-liquid state isomorphs in regard to invariance properties. This confirms that even glass configurations obey the hidden-scale-invariance condition \eq{hsi}, which is not limited to equilibrium configurations \cite{dyr20}.

\section{Summary}

This paper has studied three different compositions of the CuZr system by computer simulations using the efficient EMT potentials. We have traced out isomorphs in the liquid and glass phases of the three systems. Good isomorph invariance was observed of structure and dynamics in both phases, showing that the atoms move about each other in much the same way at state points on the same isomorph. This means that the thermodynamic phase diagram of CuZr systems is effectively one-dimensional. Thus for many purposes, in order to get an overview of the CuZr system, it is enough to investigate state points belonging to different isomorphs. It should be noted, though, that some quantities like, e.g., the bulk modulus are not isomorph invariant even when given in reduced units \cite{hey19}. On the other hand, most material quantities \textit{are} isomorph invariant in reduced units, e.g., the shear modulus, the shear viscosity, the heat conductivity, etc \cite{hey19}.

It would be interesting to confirm the above results using the EAM potentials. Given that the EAM and EMT both have been shown to nicely reproduce metal properties, we do not expect significantly different results. Indeed, the fact that isomorph theory describes metals well has been validated in DFT simulations of crystals \cite{hum15}.

\begin{acknowledgments}
	This work was supported by the VILLUM Foundation's \textit{Matter} grant (16515).
\end{acknowledgments}

\end{document}